\documentstyle[prd,aps,multicol,eqsecnum,amssymb]{revtex}
\begin{document}
\draft 

\title{$D=4$ supergravity dynamically coupled to 
a massless superparticle in a superfield Lagrangian approach}

\author{Igor A. Bandos$^{\ast,\ddagger}$, Jos\'e A. de 
Azc\'arraga$^{\ast}$, 
Jos\'e M. Izquierdo$^{\dagger}$ and 
Jerzy Lukierski$^{\star}$} 
\address{
$^{\ast}$Departamento de F\'{\i}sica Te\'orica and IFIC, 
 46100-Burjassot (Valencia), Spain} 
\address{$^{\dagger}$Departamento de F\'{\i}sica Te\'orica,
 Facultad de Ciencias, 47011-Valladolid, Spain}
\address{ $^\star$Institute for Theoretical  Physics,
pl. Maxa Borna 9, 50-204 Wroclaw, Poland}
\address{$^{\ddagger}$Institute for Theoretical Physics, NSC KIPT, 
UA61108, 
Kharkov, Ukraine} 
\date{FTUV/02-0712 IFIC/02-29, July 12, 02. v2: Oct. 28, 02. PRD {\bf 67}, 
065003-1-23 (2003)}

\maketitle 

\def\theequation{\arabic{equation}}

\begin{abstract}

We consider the interacting system of $D=4$ $N=1$ supergravity and 
the Brink--Schwarz massless superparticle as described by the sum of their  
superfield actions, and  derive the complete set of 
superfield equations of motion for the coupled dynamical system. These 
include source terms given by derivatives of a vector superfield 
current density with support on the worldline. 
This current density is constructed from the 
spin $3/2$ and  spin $2$ 
current density `prepotentials'.
We analyze the gauge symmetry of the coupled action and show that 
it is possible to fix the gauge in such a way that 
the equations of motion reduce to those of the 
supergravity--bosonic particle coupled system.

\end{abstract}


\begin{multicols}{2}

\narrowtext 

\section*{INTRODUCTION}

There has been recently a search for selfconsistent equations for 
supergravity coupled to a superbrane. 
They are needed, in particular, for  
the analysis of anomalies in M-theory  \cite{LMT} and 
in relation to the search 
\cite{RSsusy} for a supersymmetric Brane World scenario \cite{RS}.

In lower dimensions, $D=3,4$  
(and for $D=6$ using  harmonic 
superspace \cite{GIKOS}), where a superfield action for supergravity exists, 
one may develop a conventional  
approach to the supergravity--superbrane systems by using the sum of the 
superfield action of supergravity and the superbrane action. 
Such a superfield Lagrangian description of  
the low--dimensional supergravity--superbrane coupled system 
provides a possibility 
to study the structure of the 
superfield current densities of the supersymmetric extended objects, which 
might produce some insight in the search for a new 
superfield approach to higher dimensional supergravity 
in the line of Refs. \cite{11SG}.

In this paper we give 
a fully dynamical superfield 
description 
of the simplest $D=4$ $N=1$ supergravity--superparticle  
interacting system, given by the  sum of the superfield action 
for supergravity \cite{WZ78}
and the Brink--Schwarz action for the massless superparticle \cite{BS81} . 
We derive the complete set of superfield equations of motion   
and find that 
the  {\sl superfield} generalizations of the Einstein and Rarita--Schwinger 
equations 
acquire source terms. 
Both sources are determined by the action of the  
Grassmann spinor covariant derivatives on the vector superfield current density
distribution, which, in turn, is constructed from the spin $3/2$ and 
spin $2$ current `prepotentials'.

The $D=3, 4$ superfield supergravity action \cite{WZ78}
(see also \cite{OS78,Siegel79,OS80}) possesses off-shell 
supersymmetry and can be written, after integration of Grassmann 
variables, as a spacetime supergravity action 
(see, {\it e.g.}, \cite{WZ79,1001,BW,BK,SiBook})  
involving the so--called auxiliary fields (real vector and pseudoscalar 
for `minimal' supergravity, see, {\it e.g.}, \cite{OS78,Siegel79,OS80}). 
In higher dimensions, $D=10, 11$, neither the superfield action nor the set of 
auxiliary fields are known
(see, however, \cite{BR,HNP} for linearized $D=10$, $N=1$ supergravity 
and  \cite{11SG} for 
recent progress in superfield description of $D=11$ supergravity).  
For these cases 
we proposed in \cite{BAIL} 
to use the sum of the group manifold action 
for supergravity \cite{rheo} and the superbrane action as the basis for 
a Lagrangian description 
of  {\sl dynamical}  supergravity and the 
superbrane source system. 
Then
it was shown in \cite{BdAI1} that the bosonic `limit' of such a dynamical 
system, provided by the component formulation for supergravity coupled 
to the 
bosonic brane, is selfconsistent and preserves 
$1/2$ of the local supersymmetry of  `free' supergravity 
({\it cf.} \cite{RSsusy}, where  supergravity interacting with 
bosonic branes {\sl fixed} at the orbifold fixed ``points'' is considered).

The approach of \cite{BAIL} is general and
could be applied, in principle, 
to any  coupled supergravity--superbrane dynamical system 
provided that  
the group manifold approach  to the specific supergravity considered exists 
(this requires its search if  it is 
not known, {\it e.g.} for $D=10$ type 
IIA and type IIB supergravity). 
On the other hand, the results of \cite{BAIL} (see also \cite{BdAI1,BdAI2}) 
were  not quite what one would  commonly expect.  
In particular, while 
the supersymmetric generalization  of the Einstein 
equation acquired the expected source term from the super--$p$--brane, 
the superform generalization of the Rarita--Schwinger 
equation remained sourceless 
\cite{BAIL}. One might wonder whether these properties 
would be reproduced by  the conventional superfield approach to the 
dynamically interacting system. Showing that this is indeed the case 
is an additional motivation for the present study. 

In this paper we also 
analyze the gauge symmetry of the coupled action and find that 
it is possible to fix a  
gauge in which the superparticle coordinate function 
is zero, $\hat{\theta}(\tau)=0$
\footnote{This fact reflects the Goldstone nature of the superparticle 
coordinate functions 
\cite{Polchinski,Ivanov,Pashnev} and is related to the super--Higgs effect 
\cite{Volkov} (see also \cite{IK}).}, 
and that incorporates  the Wess--Zumino (WZ) gauge for  supergravity. 
We show that in this gauge 
the  equations of motion for the supergravity--superparticle 
coupled system reduce to those for the supergravity--bosonic particle 
coupled system derived 
in \cite{BdAI1} (for any dimension $D$). 
The superfield action in this gauge should also coincide 
with the action considered in \cite{BdAI1} after integration over 
the superspace Grassmann 
coordinates $\theta$ (not to be confused with the fermionic 
{\sl function} $\hat{\theta}\equiv \hat{\theta}(\tau)$) and 
elimination of the auxiliary fields by using their (purely algebraic) 
equations of 
motion. This explains the selfconsistency of the 
supergravity--bosonic particle coupled system, which was 
studied in \cite{BdAI1}.

{This paper is organized as follows}. 
The first three sections are devoted to the minimal off--shell 
formulation of simple supergravity in $D=4$, $N=1$ superspace. 
In spite of the fact that much of the material in these sections can be 
found in books \cite{1001,BW,BK,SiBook} and original articles 
\cite{Siegel79,WZ79,BZ87,BZ85}, we have found necessary to present it here in 
a unified notation. 

Specifically, we describe in Sect. I the superspace torsion 
constraints and their consequences derived  with the use of Bianchi identities,
collect them in compact differential form and present the 
expressions for the {\it l.h.s}'s of the superfield generalizations 
of the Rarita--Schwinger and Einstein equations in terms of 
the so--called main superfields and their covariant derivatives. 
In Sec. II we describe  the complete form of the Wess--Zumino gauge (fixed 
through conditions on the superfield supergravity  
{\sl potentials}, {\it i.e.}, on the  supervielbein and 
spin connection) 
and describe the 
residual gauge symmetry which preserves this 
Wess--Zumino gauge. 

In Sect. III we present the Wess--Zumino action 
for $D=4$, $N=1$ supergravity, and comment on the derivation of `free' 
superfield equations of motion. 
Sect. IV describes the $D=4$ $N=1$ Brink--Schwarz superparticle action in 
a supergravity {\sl background} ({\it i.e.} without assuming any action 
for superfield supergravity).

In Sect. V we present the coupled action for $D=4$, $N=1$ 
supergravity--superparticle interacting system and study its gauge symmetry 
(Sect. VA) which turns out to be the `direct sum' of supergravity and 
superparticle gauge symmetries. We derive the superfield 
equations of motion for the coupled system (Sect. VB) and study the 
properties of the superfield current potential and prepotentials (Sect. VC). 
We also find the superfield generalizations of the Rarita--Schwinger and 
Einstein equations, both of which contain source terms. 

In Sect. VI we show that the gauge symmetries of the coupled system 
allow one to fix a gauge in which the superparticle  
fermionic coordinate functions are set equal to zero. 
We explain why the coupled action in this gauge reduces to 
the action of component supergravity interacting with a bosonic 
particle. We show that the dynamical equations following from the 
superfield action are reduced to the equations for the  
supergravity--bosonic particle coupled system \cite{BdAI1}
in this gauge. We comment briefly on the bosonic counterpart of this gauge in 
general relativity with sources and on
the relation of these results 
with the (super) Higgs effect in the presence of superbranes, 
and conclude in Sec. VII.
   
Some technical results and additional discussion are 
given in the Appendices. 
Appendix A describes the chiral projector in $D=4$, $N=1$ superspace. 
In Appendix B
we present the complete list of manifest  local (gauge) symmetries 
of the superspace formulation of supergravity. 
We discuss  both the {\sl active} and 
{\sl passive} form of the {\sl superspace general coordinate transformations}, 
which we call {\sl general coordinate transformations} and 
{\sl superdiffeomorphisms}, respectively (see \cite{BdAI1,BdAI2}).  
Appendix C contains more details on the Wess--Zumino gauge. 
We determine there the complete set of   
residual gauge symmetries which preserve this  gauge. 
Surprisingly, by discussing all the gauge symmetries we find that 
the Wess--Zumino gauge is invariant under the active form of the 
superspace general coordinate transformations  
(in addition to the well known {\sl spacetime local} supersymmetry and 
Lorentz symmetry as well as {\sl spacetime} diffeomorphisms). 
We discuss briefly 
the r\^ole of this additional superfield  gauge invariance.  
Finally, 
Appendix D collects more details about the symmetries of the 
Brink--Schwarz superparticle action.

\section{$D=4$ $N=1$ SUPERGRAVITY IN SUPERSPACE}

\renewcommand{\theequation}{\arabic{section}.\arabic{equation}} 
\setcounter{equation}0

In this section we summarize our conventions and some known 
facts about the off--shell description of 
$D=4$, $N=1$ supergravity in superspace. 
All the formulae in this section coincide with those in 
\cite{BW} up to some signs and numerical coefficients in definitions. 
However, they are written here in a more compact differential form notation.

\subsection{Superspace  
constraints for minimal supergravity}

Let $\{ {Z}^M\} \equiv   
\{ x^\mu, \theta^{\breve{\alpha}} \}$ 
be the coordinates of 
curved $D=4$ $N=1$ superspace $\Sigma^{(4|4)}$. 
Here $\theta^{\breve\alpha}$ ($\breve{\alpha}=1,2,3,4\;$) 
are real  Grassmann coordinates  
(in flat superspace, as well as in the Wess--Zumino gauge, a Majorana spinor 
$\theta^{\underline{\alpha}}$, ${\underline{\alpha}}=1,2,3,4$). 
An unholonomic basis of the 
cotangent superspace is provided by 
the supervielbein one--forms 
\begin{eqnarray}\label{4Ea}
& E^{A} \equiv ( E^a, E^{\underline{{\alpha}}})
= ( E^a, E^{{\alpha}}, \bar{E}_{{\dot\alpha}}) 
\quad \; ;  \nonumber \\ & \qquad 
E^a= dZ^M E_M^a(Z)\; , \quad  
 \quad \nonumber \\ & 
E^{\underline{{\alpha}}} = 
dZ^M E^{\underline{{\alpha}}}_M(Z)
\quad \leftrightarrow \quad \cases{
E^{{\alpha}}= 
dZ^M E_M^{{\alpha}}(Z)\; , \cr 
 \bar{E}^{\dot\alpha}= dZ^M \bar{E}_M^{\dot\alpha}(Z)\; .} 
\end{eqnarray}
In this paper we mainly use  Weyl spinors notation 
($\alpha=1,2$, $\dot{\alpha}=1,2$), except for  Secs. II and VI, where  
Majorana spinors are used \cite{Weyl}.

An {\sl off--shell} supergravity multiplet 
can be extracted from the 
general superfields $E_M^{a}(Z)$, 
$E_M^{\underline{\alpha}}(Z)=(E_M^{{\alpha}}(Z), 
\bar{E}_{M{\dot\alpha}}(Z))$ by imposing the 
 constraints on some 
components $T_{CB}{}^A$, $R_{CD}{}^{ab}$,  
of torsion 2--forms, 
\begin{eqnarray}\label{4WTa=def} 
T^a & :=  {\cal D}E^a = 
dE^{a} - E^b \wedge w_b {}^a\; \equiv  \qquad \nonumber \\ & \qquad 
\equiv 
{1\over 2} E^B \wedge E^C T_{CB}{}^a \; ,    \\   
\label{4WTal=def}
T^{\alpha} & :=  {\cal D}E^{\alpha} = 
dE^{\alpha} - E^\beta \wedge w_\beta {}^\alpha \equiv   
\qquad \nonumber \\ & \qquad \equiv 
{1\over 2} E^B \wedge E^C T_{CB}^{\alpha} \; , 
 \\  
\label{4WTdA=def}
T^{\dot{\alpha}} & :=  {\cal D}\bar{E}^{\dot{\alpha}} 
=
d \bar{E}^{\dot{\alpha}} - \bar{E}^{\dot{\beta}} \wedge 
w_{\dot{\beta}}{}^{\dot{\alpha}} \equiv 
\qquad \nonumber \\ & \qquad \equiv 
{1\over 2} E^B \wedge E^C T_{CB}^{\dot{\alpha}} \; , 
\end{eqnarray}
and the curvature 
\begin{eqnarray}\label{4WR=def0}
R^{ab}& := dw^{ab} -w^{ac}\wedge w_c{}^b 
\equiv {1 \over 2} E^C \wedge E^D R_{DC}{}^{ab}
\; 
\end{eqnarray}
of the spin connection one--form 
$w^{ab}=dZ^M w_M{}^{ab}= - w^{ba}$ 
\begin{eqnarray}\label{4Wom=}
& w_\beta {}^\alpha= 
{1\over 4} w^{ab} (\sigma_a\tilde{\sigma}_b)_\beta {}^\alpha  
\; , \quad w_{\dot{\beta}} {}^{\dot{\alpha}}= -
{1\over 4} w^{ab} (\tilde{\sigma}_a\sigma_b)^{\dot{\alpha}}{}_{\dot{\beta}} 
\; , 
\end{eqnarray}
The constraints of minimal supergravity \cite{WZ77,OS78,BW,Siegel79}
include $T_{\alpha \dot{\beta}}{}^a= -2i\sigma^a_{\alpha \dot{\beta}}$
as well as $T_{\alpha\beta}{}^A=0=T_{\dot{\alpha}\dot{\beta}}{}^A$, 
$T_{\alpha\dot{\beta}}{}^{\dot{\gamma}}=0$,  $T_{\alpha b}{}^c=0$, 
and $R_{\alpha\dot{\beta}}{}^{ab}=0$ (or $T_{ab}{}^c=0$ as, {\it e.g.}, 
in \cite{BW}) 
\footnote{A minimal complete set of 
superspace constraints for the minimal supergravity 
multiplet \cite{auxN1} can be found, {\it e.g.}, in 
 \cite{1001,BW,BZ87}; see \cite{Siegel79,1001,Zima,BZ87} and refs. therein for 
nonminimal supergravity multiplets, and 
\cite{AlgC} for a discussion of the algebraic origin of the supergravity 
constraints.}.
In the presence of the complete set of constraints, the Bianchi identities
\begin{eqnarray}\label{BIT} 
& {\cal D}T^A\equiv -E^B \wedge R_{B}{}^A\;  \; \Leftrightarrow \; 
\cases{ {\cal D}T^a\equiv -E^b \wedge R_{b}{}^a\; ,\cr 
 {\cal D}T^{\alpha} \equiv -E^{\beta} \wedge R_{\beta}{}^{\alpha} \; ,\cr
{\cal D}T^{\dot\alpha} \equiv -E^{\dot\beta} \wedge 
R_{\dot\beta}{}^{\dot\alpha} \; , 
\cr } 
\\  \label{BIR}
& {\cal D} R^{ab}\equiv 0\;  \;    
\Rightarrow \; \cases{ {\cal D} R_{\alpha}{}^{\beta}\equiv 0\; , \cr    
{\cal D} R_{\dot{\alpha}}{}^{\dot{\beta}}\equiv 0\; ,} 
\end{eqnarray} 
(integrability conditions for 
Eqs. (\ref{4WTa=def})--(\ref{4WR=def0}))
express the superspace torsion and curvature through the  
set of `main superfields' 
\begin{eqnarray}\label{4WGa}
& G_a:= 2i (T_{a\beta}{}^\beta -T_{a\dot{\beta}}{}^{\dot{\beta}})\; , 
\\ \label{4WR} 
& \bar{R}:= -{1\over 3} R_{{\alpha}{\beta}}{}^{{\alpha}{\beta}} = (R)^* \; , 
\qquad \\ 
\label{4WchW}
& W^{\alpha\beta\gamma} := 
4i \tilde{\sigma}^{c\dot{\gamma}\gamma} R_{\dot{\gamma}c}{}^{\alpha\beta}
= W^{(\alpha\beta\gamma)}
= (\bar{W}^{\dot{\alpha}\dot{\beta}\dot{\gamma}})^* \; . 
\end{eqnarray}

The constraints of minimal supergravity and their consequences 
can be collected in the following expressions for the superspace torsion 
2--forms ({\it cf.} \cite{BW})
\begin{eqnarray}\label{4WTa=} 
T^a & 
=- 2i\sigma^a_{\alpha\dot{\alpha}} E^\alpha \wedge \bar{E}^{\dot{\alpha}} 
+{1\over 16} E^b \wedge E^c \varepsilon^a{}_{bcd} 
G^d \; ,  
\\ 
\label{4WTal=}
T^{\alpha} & 
= {i\over 8} E^c \wedge E^{\beta} (\sigma_c\tilde{\sigma}_d)_{\beta}
{}^{\alpha} G^d  - \hspace{2.3cm} 
\nonumber \\ &  -{i\over 8} E^c \wedge \bar{E}^{\dot{\beta}} 
\epsilon^{\alpha\beta}\sigma_{c\beta\dot{\beta}}R 
+
 {1\over 2} E^c \wedge E^b \; T_{bc}{}^{\alpha}\; , \\ 
\label{4WTdA=}
T^{\dot{\alpha}} 
& = {i\over 8} E^c \wedge E^{\beta} \epsilon^{\dot{\alpha}\dot{\beta}}
\sigma_{c\beta\dot{\beta}}
\bar{R} - \hspace{2.3cm}  
\;   
\nonumber \\ &   
-{i\over 8} E^c \wedge \bar{E}^{\dot{\beta}} 
(\tilde{\sigma}_d\sigma_c)^{\dot{\alpha}}{}_{\dot{\beta}}
\, G^d +
 {1\over 2} E^c \wedge E^b \; T_{bc}{}^{\dot{\alpha}}\; .
\end{eqnarray}

The superspace Riemann curvature 2--form  
is determined by 
\begin{eqnarray}\label{4WR=def}
R^{ab}& := dw^{ab} -w^{ac}\wedge w_c{}^b = \qquad {} \qquad 
{} \quad \nonumber \\ & = {1\over 2} R^{\alpha\beta} 
(\sigma^a\tilde{\sigma}^b)_{\alpha\beta} 
- {1\over 2} R^{\dot{\alpha}\dot{\beta}} 
(\tilde{\sigma}^a\sigma^b)_{\dot{\alpha}\dot{\beta}} \; , 
\end{eqnarray}
with 
\begin{eqnarray}\label{4WR=}
R^{\alpha\beta} & \equiv  dw^{\alpha\beta} - w^{\alpha\gamma} 
\wedge w_\gamma{}^\beta \equiv {1 \over 4} 
R^{ab} (\sigma_a\tilde{\sigma}_b)^{\alpha\beta}= \nonumber
\\ 
& 
= -{1\over 2} E^\alpha \wedge E^\beta \bar{R} 
-{i\over 8} E^c \wedge E^{(\alpha}\,  
\tilde{\sigma}_c{}^{\dot{\gamma}\beta)} \bar{{\cal D}}_{\dot{\gamma}}\bar{R}+ 
\nonumber \\ 
& 
+{i\over 8} E^c \wedge E^{\gamma} 
(\sigma_c\tilde{\sigma}_d)_{\gamma}{}^{(\beta} {\cal D}^{\alpha)} G^d -  
\nonumber \\ & 
-{i\over 8}  E^c \wedge \bar{E}^{\dot{\beta}} \sigma_{c\gamma\dot{\beta}} 
W^{\alpha\beta\gamma} + {1\over 2} E^d \wedge E^c R_{cd}{}^{\alpha\beta} \; ,
\end{eqnarray}
and $R^{\dot{\alpha}\dot{\beta}}= 
(R^{{\alpha}{\beta}})^*$. 

Note that in our conventions 
the spinor covariant derivatives ${{\cal D}}_{{\alpha}}= 
- (\bar{{\cal D}}_{\dot{\alpha}})^*$ 
are defined by 
the following decomposition of 
the covariant differential 
${\cal D}$ 
\begin{eqnarray}\label{4WD}
{\cal D}:= & E^A {\cal D}_A = E^a {\cal D}_a +  E^{\underline{\alpha}} 
{\cal D}_{\underline{\alpha}}= 
\nonumber \\
& =E^a {\cal D}_a +  E^{\alpha} {\cal D}_{\alpha}+ 
\bar{E}^{\dot{\alpha}} \bar{{\cal D}}_{\dot{\alpha}} \; . 
\end{eqnarray}
(hence, 
${{\cal D}}_{\underline{{\alpha}}}=
({\cal D}_{{\alpha}}\, , \,   
- \bar{{\cal D}}^{\dot{\alpha}})$;  note the minus sign).
Then, since {\it e.g.}, ${\cal D}_\alpha= E_\alpha{}^M\partial_M +
w_\alpha$, it is also natural that 
the spinor components of the spin connection form  
$w^{ab}= dZ^M w_M^{ab}= E^A w_A^{ab}:=
E^c w_c^{ab}+ E^{\alpha} w_{\alpha}{}^{ab} +  \bar{E}^{\dot{\alpha}}
w_{\dot{\alpha}}{}^{ab}$ be related by 
$w_{\alpha}{}^{ab}= -(w_{\dot{\alpha}}{}^{ab})^*$ 
(hence, 
$w_{\underline\alpha}{}^{ab}=(w_{\alpha}{}^{ab}, 
- w^{\dot{\alpha}}{}^{ab})$).  

The Bianchi identities (\ref{BIT}), (\ref{BIR}) imply as well that 
the main superfields (\ref{4WGa}), (\ref{4WR}), (\ref{4WchW})  
obey the equations
\begin{eqnarray}
\label{chR}
& {\cal D}_\alpha \bar{R}=0\; , \qquad \bar{{\cal D}}_{\dot{\alpha}} {R}=0\; ,
 \\ 
\label{chW} & \bar{{\cal D}}_{\dot{\alpha}} W^{\alpha\beta\gamma}= 0\; , 
\qquad 
{{\cal D}}_{{\alpha}} \bar{W}^{\dot{\alpha}\dot{\beta}\dot{\gamma}}= 0\;,
\\ 
\label{DG=DR} &
\bar{{\cal D}}^{\dot{\alpha}}G_{{\alpha}\dot{\alpha}}= {\cal D}_{\alpha} R 
\; , \qquad 
{{\cal D}}^{{\alpha}}G_{{\alpha}\dot{\alpha}}= 
\bar{{\cal D}}_{\dot{\alpha}} \bar{R} \; , 
\\ \label{DW=DG} &
{{\cal D}}_{{\gamma}}W^{{\alpha}{\beta}{\gamma}}= 
\bar{{\cal D}}_{\dot{\gamma}} {{\cal D}}^{({\alpha}}G^{{\beta})\dot{\gamma}}
\; , \qquad \nonumber \\ & 
\bar{{\cal D}}_{\dot{\gamma}} \bar{W}^{\dot{\alpha}\dot{\beta}\dot{\gamma}}
= {{\cal D}}_{{\gamma}} \bar{{\cal D}}^{(\dot{\alpha}|} 
G^{{\gamma}|\dot{\beta})} \; . \qquad
\end{eqnarray}

For the sake of brevity, we will call `constraints' the complete set 
of relations 
(\ref{4WTa=})--(\ref{4WTdA=}), 
(\ref{4WR=}), 
(\ref{chR})--(\ref{DW=DG}).

\subsection{Off--shell nature of the constraints}

Using the Bianchi identities (\ref{BIT}), (\ref{BIR}), 
one also finds  that 
the fermionic torsion  components $T_{ab}{}^{\alpha}$ 
$T_{ab}{}^{\dot{\alpha}}$ entering  
Eqs. (\ref{4WTal=}), (\ref{4WTdA=}) 
(which may be regarded as superfield generalizations of the
gravitino field strengths, see 
below (\ref{TbbfWZ}))
are also expressed through    
the main superfields (\ref{4WGa}), (\ref{4WR}), (\ref{4WchW})
\begin{eqnarray}\label{Tabg}
T_{{\alpha}\dot{\alpha}\; \beta \dot{\beta }\; {\gamma}} 
& \equiv   \sigma^a_{{\alpha}\dot{\alpha}} 
\sigma^b_{\beta \dot{\beta }} 
\epsilon_{{\gamma}{\delta}} T_{ab}{}^{{\delta}}=  
-{1\over 8}  \epsilon_{{\alpha}{\beta}} {\bar{{\cal D}}}_{(\dot{\alpha}|}
G_{\gamma |\dot{\beta})} -  
\nonumber 
\\ 
& -{1\over 8} \epsilon_{\dot{\alpha}\dot{\beta}}[W_{\alpha \beta\gamma} - 
2\epsilon_{\gamma (\alpha}{\cal D}_{\beta)} R] 
 \; , 
\\ \label{Tabdg}
T_{{\alpha}\dot{\alpha}\; \beta \dot{\beta }\; \dot{\gamma}} 
& \equiv  \sigma^a_{{\alpha}\dot{\alpha}} 
\sigma^b_{\beta \dot{\beta }} 
\epsilon_{\dot{\gamma}\dot{\delta}} T_{ab}{}^{\dot{\delta}} = 
{1 \over 8} \epsilon_{\dot{\alpha}\dot{\beta}} 
{{\cal D}}_{({\alpha}}
G_{{\beta})\dot{\gamma}} - 
  \nonumber \\ 
& -{1 \over 8} \epsilon_{{\alpha}{\beta}} 
[ \bar{W}_{\dot{\alpha} \dot{\beta}\dot{\gamma}} 
+ 2 \epsilon_{\dot{\gamma}(\dot{\alpha}}  {\bar{{\cal D}}}_{\dot{\beta})}
\bar{R}] \; .
\end{eqnarray}
Eqs. (\ref{Tabg}), (\ref{Tabdg}) imply, in particular, 
\begin{eqnarray}\label{gr1}
& (\sigma^a \tilde{\sigma}^b)_\beta{}^\gamma T_{ab\gamma}= 
{3\over 4} {\cal D}_{\beta} R\; , \nonumber \\ &  
(\tilde{\sigma}^a\sigma^b)^{\dot{\gamma}}{}_{\dot{\beta}} 
T_{ab\dot{\gamma}}= {3\over 4} \bar{\cal D}_{\dot{\beta}} \bar{R}
\; .
\end{eqnarray}

Moreover, the {\it l.h.s} of the Rarita--Schwinger equation can be identified 
with the leading component 
({\it i.e.}, the $\theta =0$ value) of the superfield expression 
$\epsilon^{abcd}T_{bc}{}^{\alpha}\sigma_{d\alpha\dot{\alpha}}$ 
(see (\ref{SGRS}) and {\it e.g.}, \cite{BW}). 
Using the Pauli matrix algebra ($\sigma^a\tilde{\sigma}^{b}= 
\eta^{ab} I + {i\over 2} \epsilon^{abcd} 
\sigma_{c}\tilde{\sigma}_d$, $\sigma^{[a}\tilde{\sigma}^{b]}{\sigma}_{b}=
3{\sigma}^{a}$) one finds from Eq. (\ref{Tabg}) 
\begin{eqnarray}\label{SGRS=off}
\Psi^a_{\dot{\alpha}} & := 
\epsilon^{abcd}T_{bc}{}^{\alpha}\sigma_{d\alpha\dot{\alpha}}= \hspace{2cm} 
\nonumber \\
& ={i\over 8} \tilde{\sigma}^{a\dot{\beta}\beta} \bar{{\cal D}}_{(\dot{\beta}|}
G_{\beta|\dot{\alpha})} + 
{3i\over 8} {\sigma}^a_{\beta \dot{\alpha}} {\cal D}^{\beta}R \; .
\end{eqnarray}

The {\sl fields}
${\cal D}^{\beta}R \vert_{\theta=0}$ and  
$\bar{{\cal D}}_{(\dot{\beta}|}
G_{\beta|\dot{\alpha})} \vert_{\theta=0}$ are not restricted by the 
constraints (\ref{4WTa=})--(\ref{4WR=}), (\ref{chR})--(\ref{DW=DG}). 
They are  arbitrary fermionic functions, 
which rather can be  identified with the 
corresponding irreducible parts of the 
leading component $\Psi^a_{\dot{\alpha}}\vert_{\theta=0}$ 
of the Rarita--Schwinger superfield  $\Psi^a_{\dot{\alpha}}$. 
Hence, $\Psi^a_{\dot{\alpha}}\vert_{\theta=0}$ 
remains arbitrary. 

Similarly, 
the bosonic Riemann curvature tensor superfield 
is determined by 
\begin{eqnarray}\label{Rab}
& \sigma^c_{\gamma\dot{\gamma}}\sigma^d_{\delta\dot{\delta}}
R_{cd}{}^{\alpha\beta}= - 2 \epsilon_{{\gamma}{\delta}}
r_{\dot{\gamma}\dot{\delta}}{}^{\alpha\beta} 
-2 \epsilon_{\dot{\gamma}\dot{\delta}} 
r_{{\gamma}{\delta}}{}^{\alpha\beta} \\ 
\label{rab} & 
r_{\dot{\gamma}\dot{\delta}}{}^{\alpha\beta} = {1\over 16} 
\bar{{\cal D}}_{(\dot{\gamma}} 
{{\cal D}}^{({\alpha}} G^{{\beta})}{}_{\dot{\delta})} 
\\ 
\label{brab} & 
r_{{\gamma}{\delta}}{}^{\alpha\beta}= -  {1\over 16} 
{{\cal D}}_{({\gamma}} W_{{\delta})}{}^{\alpha\beta} 
-  {1\over 32} \delta_{({\gamma}}^{\alpha}\delta_{{\delta})}^{\beta}
(\bar{{\cal D}}\bar{{\cal D}}\bar{R}- 2 R\bar{R})\; . 
\end{eqnarray}
In particular, (\ref{brab}) indicates that the superfield generalization 
of the (spin--tensor components 
of the) Weyl tensor, $C_{\alpha\beta{\gamma}{\delta}}= 
C_{(\alpha\beta{\gamma}{\delta})}$, are defined through 
the nonvanishing spinor derivative of ${W}_{{\alpha}{\beta}{\gamma}}$ 
\begin{eqnarray}
\label{Weyl} & 
C_{\alpha\beta{\gamma}{\delta}}:= r_{(\alpha\beta{\gamma}{\delta})}
= -  {1\over 16} {{\cal D}}_{({\alpha}} W_{\beta{\gamma}{\delta})} \; .
\end{eqnarray}
In this sense one says that 
${W}_{{\alpha}{\beta}{\gamma}}$ and its complex 
conjugate $\bar{W}_{\dot{\alpha} \dot{\beta}\dot{\gamma}}$ 
provide a superfield 
generalization of the Weyl tensor.  

The superfield generalization of the Ricci tensor
is given by 
\begin{eqnarray}
\label{RRici}  
R_{bc}{}^{ac}& = {1\over 32} ({{\cal D}}^{{\beta}}
\bar{{\cal D}}^{(\dot{\alpha}|} G^{{\alpha}|\dot{\beta})} -
\bar{{\cal D}}^{\dot{\beta}} {{\cal D}}^{({\beta}}G^{{\alpha})\dot{\alpha}})
\sigma^a_{\alpha\dot{\alpha}}\sigma_{b\beta\dot{\beta}} 
- \hspace{-1cm}
\nonumber \\ & - {3\over 64} (\bar{{\cal D}}\bar{{\cal D}}\bar{R}
+ {{\cal D}}{{\cal D}}{R}- 4 R\bar{R})\delta_b^a\; , 
\end{eqnarray}
and, henceforth, the scalar curvature superfield is 
\begin{eqnarray}
\label{Rscalar}  
R_{ab}{}^{ab}= 
- {3\over 16} (\bar{{\cal D}}\bar{{\cal D}}\bar{R}
+ {{\cal D}}{{\cal D}}{R}- 4 R\bar{R})\; .  
\end{eqnarray}

Hence, once again, one can identify the (arbitrary) leading 
components of the corresponding  second derivatives of main 
superfields $G_a$ and $R$ (entering the {\it r.h.s.} of Eq. (\ref{RRici}))   
with the irreducible components of the 
Ricci tensor $R_{bc}{}^{ac}\vert_{\theta=0}$
(or Einstein tensor $(R_{bc}{}^{ac}-{1\over 2}
\delta_b{}^aR_{dc}{}^{dc} )\vert_{\theta=0}$) which, thus, 
remains arbitrary after imposing the 
constraints (\ref{4WTa=})--(\ref{4WR=}). 

This exhibits the well known 
fact that the constraints (\ref{4WTa=})--(\ref{4WR=}) describe the 
{\sl off--shell} supergravity multiplet.

\section{Wess--Zumino (WZ) gauge}

To move from the superfield formulation of supergravity 
to the component formulation ({\it i.e.} in terms of 
spacetime fields) \cite{WZ79,BW}, 
one fixes the so--called Wess--Zumino (WZ) gauge, where, {\sl in particular}, 
\footnote{
We mainly use in Sec. II Majorana spinor notations    
$E^{\underline{\alpha}}=(E^{\alpha}, E_{\dot{\alpha}})$ 
\cite{Weyl}; 
this also makes all the formulae 
of this section, except Eq. (\ref{TbbfWZ}), 
applicable to any dimension $D$.}
 \begin{eqnarray}
\label{WZgEff} 
 E_{\breve\alpha}{}^a\vert_{\theta =0} = 0 \; , \quad 
 E_{\breve\alpha}{}^{\underline{\beta}}\vert_{\theta =0} = 
\delta_{\breve\alpha}{}^{\underline{\beta}} \; , \quad 
w_{\breve\alpha}^{ab}\vert_{\theta =0} = 0 \; , 
\end{eqnarray}
while 
\begin{eqnarray}\label{WZ0gg} 
E_\mu^{\; a}\vert_{\theta =0} = e_\mu^{\; a}(x) \; , \qquad 
E_\mu^{\; \underline{\alpha}}\vert_{\theta =0} =
\psi_\mu^{\underline{\alpha}}\, , 
\\ \label{WZ0w} 
w_\mu^{ab}\vert_{\theta =0} = \omega_\mu^{ab}(x)
\end{eqnarray}
remain unrestricted and are identified with the vielbein, gravitino and 
(composed) spin--connection fields of the component formulation of 
supergravity \cite{WZ79,BW}.

One can collect the expressions for the supervielbein superfield 
in (\ref{WZgEff}),  (\ref{WZ0gg}) 
in the matrix relation
\begin{eqnarray}\label{WZ0gg1} 
E_N{}^A\vert_{\theta =0} = \left(\matrix{e_\nu^a(x) & 
\psi_\nu^{\underline{\alpha}}(x)\cr 
0 & \delta_{\breve{\beta}}{}^{\underline{\alpha}} }\right) \; . 
\end{eqnarray}
Their evident consequences are 
\begin{eqnarray}\label{WZ0gg2} 
E_A{}^N\vert_{\theta =0} = \left(\matrix{e_a^\nu(x) & 
- \psi_a^{\breve{\beta}}(x)\cr 
0 & \delta_{\underline{\alpha}}{}^{\breve{\beta}}}\right) \; ,  
\end{eqnarray}
where $\psi_a^{\breve{\beta}}(x)\equiv 
e_a^\nu \psi_\nu^{\underline{\alpha}}(x)
\delta_{\underline{\alpha}}{}^{\breve{\beta}}$. 

Note that already these simple formulae allow one to derive, 
{\it e.g.}, the following useful formula 
\begin{eqnarray}\label{TbbfWZ} 
T_{ab}{}^{\alpha}\vert_{\theta =0} & = 2 e_a^\mu e_b^\nu 
{\cal D}_{[\mu}\psi_{\nu]}^{\alpha}(x) - \hspace{2.5cm} \nonumber \\ 
& - {i\over 4} (\psi_{[a}\sigma_{b]})_{\dot{\beta}}
G^{\alpha\dot{\beta}}\vert_{\theta =0} 
 - {i\over 4} (\tilde{\sigma}_{[a}\bar{\psi}_{b]})^{\alpha} 
R\vert_{\theta =0} \; ,
\end{eqnarray}
where ${\cal D}_{[\mu}\psi_{\nu]}^{\alpha}(x)= 
\partial_{[\mu} \psi_{\nu]}^{\alpha}(x) - 
\psi_{[\nu}^{\beta}(x) \, w_{\mu]\beta}{}^{\alpha}\vert_{\theta =0}$ 
is the gravitino fields strength (though with the nonstandard spin connection 
which, in general, due to (\ref{4WTa=}), involves 
the term proportional  to $G_a\vert_{\theta =0}$ into the spacetime 
torsion). 
Thus one can call $T_{ab}{}^{\alpha}$ the superfield generalization 
of the gravitino field strength.

One more simple but useful equation which is valid 
due to Eqs. (\ref{WZgEff}), (\ref{WZ0gg})  is
\begin{eqnarray}\label{Ber0=e}
E\vert_{\theta=\bar{\theta}=0}\equiv sdet(E_M^A(x,0,0))= det (e_\mu^a)
\equiv e(x) \; . 
\end{eqnarray}

\subsection{Complete description  of the Wess--Zumino gauge}

As it was early recognized \cite{BZ85,NormalC,NormalC11},  the WZ
gauge is the fermionic counterpart of the normal coordinate 
system in General Relativity (see refs. in \cite{OS80,NormalS} 
and \cite{OS80,NormalS,BZ85} for the so--called 
normal gauge in supergravity, which is the complete 
superspace generalization of the normal coordinate frame). 
This observation suggested to collect \cite{BZ85} the complete set of the 
conditions of the WZ 
gauge in 
\footnote{
Note that there exists another ({\sl `prepotential'}) form of the 
Wess--Zumino gauge which is fixed through a condition for 
the Ogievetsky--Sokatchev auxiliary vector {\sl prepotential},  
giving ${\cal H}^\mu = 
\theta\sigma^a\bar{\theta} e_a^\mu(x) + \bar{\theta}\bar{\theta} 
\theta^\alpha \psi_\alpha^\mu (x) + c.c. +  {\theta}{\theta} \, 
\bar{\theta}\bar{\theta} \, A^\mu(x) $ \cite{OS78}, 
and for the chiral compensator,  $\Phi= e^{1/3}(1 - 
{2\over 3} \theta\sigma^a\bar{\psi}_a + \ldots)$ (see, e.g.,  
\cite{Siegel79,BZ85}).} 
\begin{eqnarray}\label{WZgauge}
\theta^{\breve\alpha} E_{\breve\alpha}^{\, a}=0\; , \qquad  
\theta^{\breve\alpha} (E_{\breve\alpha}^{\, \underline{\beta}}- 
\delta_{\breve\alpha}^{\, \underline{\beta}}) 
=0\; , \\ \nonumber 
\theta^{\breve\alpha} w_{\breve\alpha}^{ab}=0\; . 
\end{eqnarray}
Using the inner product notation (see Eqs. 
(\ref{itEA}), (\ref{itwab})), the WZ gauge may be equivalently defined by  
\begin{eqnarray}\label{iWZg}
i_\theta E^{a}=0\; , \quad \\  \label{iWZgf}
i_\theta E^{\underline{\alpha}}= \theta^{\breve\beta} 
\delta_{\breve\beta}^{\, \underline{\alpha}}\equiv \theta^{\underline{\alpha}}
\; , \\ \label{iWZw}
i_\theta w^{ab}=0\; , 
\end{eqnarray}
where $\theta^{\underline{\alpha}}$ is a 
Grassmann coordinate 
with a tangent space spinor index,   
\begin{eqnarray}\label{WZgth}
& \theta^{\underline{\beta}} \equiv \theta^{\breve\alpha}  
\delta_{\breve\alpha} {}^{\underline{\beta}} \; .    \qquad 
\end{eqnarray}

One of the characteristic properties of the WZ 
gauge 
(\ref{WZgauge}) is that the Grassmann coordinate (\ref{WZgth})
coincides  with the contraction of the fermionic supervielbein form, 
(\ref{iWZgf}). 
The next observation is that in the gauge (\ref{WZgauge})  
\begin{eqnarray}
\label{WZgD}
& \theta^{\breve\alpha} {\cal D}_{\breve\alpha} = 
\theta^{\underline{\beta}} \; {\cal D}_{\underline{\beta}} = 
 \theta^{\breve\alpha} \partial _{\breve\alpha} \equiv \theta\partial\; . 
\end{eqnarray} 
With this in mind one can find that the decomposition of 
the supervielbein and spin connection superfields
can be expressed in terms of the physical graviton and gravitino fields  
(Eq. (\ref{WZ0gg})), the leading components of the torsion and curvature 
superfields and their covariant derivatives. A convenient way of 
reproducing these decompositions is by using  the following 
recurrent relations ({\it cf.} \cite{BZ85})
\begin{eqnarray} 
\label{WZg1b} 
& (1+ \theta\partial) E^a = i_\theta T^a + dx^\mu E_\mu^a \; 
\\ 
\label{WZg1f}
& (1+ \theta\partial) E^{\underline{\alpha}} = 
{\cal D}\theta^{\underline{\alpha}}+ i_\theta T^{\underline{\alpha}} + 
dx^\mu E_\mu^{\underline{\alpha}}\; , 
\\ \label{WZg1w} 
& (1+\theta \partial )w^{ab} =i_\theta R^{ab}+ 
dx^\mu w_\mu^{ab}
\; , 
\end{eqnarray}
together with Eq. (\ref{WZgD}). 
Eqs. (\ref{WZg1b}), (\ref{WZg1f}), 
(\ref{WZg1w}) are obtained by    
taking the external derivative of  the defining relations of the WZ 
gauge, Eqs. (\ref{WZgauge}). There 
 \begin{eqnarray}\label{WZgDth0}
& {\cal D} \theta^{\underline{\beta}} =   
d\theta^{\underline{\beta}} -  
\theta^{\underline{\gamma}} w_{\underline{\gamma}}{}^{\underline{\beta}}
\; , \qquad 
\\ \label{WZgitT}
& i_\theta T^{A} \equiv E^{C} \theta^{\underline{\beta}} 
T_{{\underline{\beta}}C}{}^{A}\; , \qquad 
i_\theta R^{ab} \equiv E^{D} \theta^{\underline{\gamma}} 
R_{{\underline{\gamma}}{D}}{}^{ab}\; .
\end{eqnarray}
Eqs. (\ref{WZg1b}), (\ref{WZg1f}), (\ref{WZg1w}) 
do not restrict the physical fields (\ref{WZ0gg}), as 
the terms containing $dx^\mu$ in {\it l.h.s}'s are canceled 
by the last terms in the {\it r.h.s}'s. 
Thus the leading ($\theta=0$) components of Eqs. (\ref{WZg1b}), (\ref{WZg1f}), 
(\ref{WZg1w})  reproduce Eqs. (\ref{WZgEff}). 

A discussion of the decomposition 
of the superfields, {\it i.e.} of the solutions to Eqs. 
(\ref{WZg1b}), (\ref{WZg1f}) and 
(\ref{WZg1w}), can be found in Appendix C1.

\subsection{Symmetries preserving the Wess--Zumino gauge}

In the consideration of the coupled system it is important to know 
the subset of superspace local symmetries preserving the gauge 
(\ref{WZgauge}). 
The subset of superspace diffeomorphisms (parameter $b^M(Z)$) 
and local Lorentz ($L^{ab}(Z)$) transformations preserving 
this gauge is singled out by the equations (see Appendix C2 
for details and further discussion) 
\begin{eqnarray} 
\label{WZgpr1} 
& \theta \partial(b^A) = (b^B) 
\theta^{\underline{\gamma}} T_{\underline{\gamma}B}{}^A 
+ \qquad \nonumber \\ & \qquad 
+ \theta^{\underline{\gamma}} (L_{\underline{\gamma}}{}^{\underline{\beta}}
- b^M w_{M\underline{\gamma}}{}^{\underline{\beta}}) 
\delta_{\underline{\beta}}{}^A
\; , 
\\ \label{WZgpr2} 
&  \theta \partial(L^{ab}(Z)- b^M w_M{}^{ab})=  
- b^D \theta^{\underline{\gamma}}  
R_{\underline{\gamma}D}{}^{ab} \; , 
\end{eqnarray} 
where 
\begin{eqnarray}
\label{Ldec}
& L_B{}^A (Z)= \left(\matrix{L_b{}^a & 0 \cr 
              0 & L_{\underline{\beta}}{}^{\underline{\alpha}}}\right)\; ,
\quad L^{ab}= - L^{ba}\; , \nonumber \\ & 
L_{\underline{\beta}}{}^{\underline{\alpha}}= 
{1\over 4} L^{ab} \gamma_{ab}{}_{\underline{\beta}}{}^{\underline{\alpha}}\; , 
\end{eqnarray}
and the parameter
$b^M$ can be conventionally decomposed into a 
fermionic spinor and a bosonic vector part 
\begin{eqnarray}
 \label{bdecWZ}
& b^A := b^M(Z) E_M^A(Z)\equiv (b^a(Z), 
\varepsilon^{\underline{\alpha}} (Z))\; , 
\end{eqnarray}

It is instructive to write Eqs.  (\ref{WZgpr1}),  (\ref{WZgpr2}) 
in the weak field approximation. 
At zero--order 
one finds the set of equations
\begin{eqnarray}
\label{WZgrs1} 
& \theta \partial(b^a) = - 2i 
\varepsilon^{\underline{\beta}} 
\gamma^a_{\underline{\beta}\underline{\gamma}}
\theta^{\underline{\gamma}} \; , 
\\ \label{WZgrs2} 
&  \theta \partial(\varepsilon^{\underline{\alpha}})=
\theta^{\underline{\beta}} L_{\underline{\beta}}{}^{\underline{\alpha}}\; , 
\\ \label{WZgrs3} 
&  \theta \partial L^{ab}(Z)= 0\; , 
\end{eqnarray}
which can be easily solved, 
\begin{eqnarray}
\label{WZgrs1s} 
& b^a(Z)  = b_0^a(x) + 2i \theta\gamma^a\varepsilon_0(x) + 
{i \over 4} \theta (\gamma_{bc}\gamma^a)\theta \, l^{bc}(x) \; , 
\\ \label{WZgrs2s} 
& \varepsilon^{\underline{\alpha}}(Z)
= 
\varepsilon_0^{\underline{\alpha}}(x) - \theta^{\underline{\beta}}
l_{\underline{\beta}}{}^{\underline{\alpha}}(x)  \; ,
\hspace{1.3cm}
\\ \label{WZgrs3s} 
& L^{ab}(Z) = l^{ab}(x)\; ,  \hspace{2.3cm}
\end{eqnarray}
where $b_0^a(x)$, $\varepsilon_0^{\underline{\alpha}}(x)$ are arbitrary 
vector and spinor functions and $l^{ab}(x)$ are local Lorentz parameters.

In the general case the WZ 
gauge is also preserved, in particular (see Appendix C2) 
by {\sl spacetime} diffeomorphisms (with parameters 
$b^a(Z)\vert_{\theta=0}$), 
as well as by 
Lorentz ($l^{ab}(x)$) and local supersymmetry 
($\varepsilon^{\underline{\alpha}}(x)= \varepsilon^{\underline{\alpha}}(Z)
\vert_{\theta=0}$) transformations.

\section{SUPERFIELD ACTION FOR `FREE' $D=4$ $N=1$ SUPERGRAVITY}

\renewcommand{\theequation}{\arabic{section}.\arabic{equation}} 
\setcounter{equation}0

\subsection{Superfield action and variational problem with constraints}

The $D=4$ $N=1$ supergravity action can be written \cite{WZ78} 
as an integral over superspace $\Sigma^{(4|4)}$ of the 
Berezenian (superdeterminant) 
$E:= sdet(E_M^A)$ of the supervielbein $E_M^A(Z)$,  
\begin{eqnarray}\label{SGact}
S_{SG} = \int d^4 x \tilde{d}^4\theta \; sdet(E_M^A) \; 
\equiv \int d^8Z \; E \; , 
\end{eqnarray} 
where $E_M^A(Z)$ are assumed to be subject to the constraints 
(\ref{4WTa=}), (\ref{4WTal=}), (\ref{4WTdA=}), (\ref{4WR=}). 
This action is evidently 
invariant under the superdiffeomorphisms and local Lorentz symmetries 
(further discussion of its gauge symmetries can be found in Appendix B1).  

One of the ways 
to obtain the superfield equations of motion from this action  
is to solve the constraints in terms of unconstrained 
superfields (prepotentials): axial vector superfield 
${\cal H}^{\mu}(x,\theta)$ \cite{OS78} and chiral compensator 
$\Phi$ \cite{Siegel79} (in this way the local symmetries of 
the complete 
superfield formulation are partially gauge fixed). 

Alternatively, following \cite{WZ78}, one can keep  $E_M^A(Z)$ as the 
basic variable, but take the constraints into account when searching for 
the independent variations. 
Namely, one denotes   
the general variation of the supervielbein and spin connections by 
\cite{WZ78}
\begin{eqnarray}\label{varEMA}
\delta E_M^{\, A}(Z) = E_M^{\, B} {\cal K}_{B}^{\, A} 
(\delta )\; , \quad 
\delta w_M^{ab}(Z) = E_M^{\, C} u_{{C}}^{ab} (\delta )\; , 
\end{eqnarray}
and obtains the  equations to be satisfied by 
${\cal K}_{B}^{\, A}  (\delta )$, 
$u_{C}^{ab} (\delta )$ from the requirement 
that the constraints (\ref{4WTa=}), (\ref{4WTal=}) are preserved 
under (\ref{varEMA}). Then one  
solves  these equations in terms of some set of  independent variations. 
Straightforward but quite involved calculations 
(the results of which were partially given in \cite{WZ78}) 
show that 
the constraints of minimal supergravity 
(\ref{4WTa=})--(\ref{4WR=}) are preserved by a set of {\sl superfield 
variations} (superspace coordinates are not affected) which include: 
\\ 
{\it i)} the local Lorentz transformations 
${\delta}_{L}(L^{ab})$, Eq. (\ref{LorentzS}); \\ 
{\it ii)} 
the variational version of the superspace general coordinate 
transformations \cite{WZ78} ($\tilde{\delta}_{gc}(t^A)$, Eqs. 
(\ref{tgcZ}), (\ref{tgcE}), (\ref{tgcwab}) in Appendix B); \\ 
{\it iii)} the set of transformations with  parameters 
$\delta H^a= 
{1\over 2} \sigma^a_{\alpha\dot\alpha} \delta H^{\alpha\dot\alpha}$,  
$\delta {\cal U}$, $\delta {\bar{\cal U}}$, 
under which the supervielbein transforms as 
\footnote{This procedure can be regarded as a linearized counterpart 
of solving the superspace constraints in terms of the prepotentials 
\cite{Siegel79} (the price to achieve linearity, however, is 
that we have to deal 
with the  covariant derivatives
${\cal D}_A$ rather than with the holonomic ones, $\partial_M$). 
So, the counterpart $\delta H^a$ of the variation of the 
Ogievetsky--Sokatchev auxiliary vector superfield \cite{OS78} 
${\cal H}^\mu$, as well as the counterparts of the variation of 
the complex chiral compensators $\Phi$ \cite{Siegel79}, 
$({\cal D}{\cal D}- \bar{R})\delta {\cal U}$, 
are involved in the solution of these equations.  
[The (anti)chiral superfield $\bar{\Phi}$ 
satisfies ${{\cal D}}_{{\alpha}}\bar{\Phi}=0$ 
and can be expressed through the independent 
superfield ${{\cal U}}$ by $\bar{\Phi}= 
({{\cal D}}{{\cal D}}- \bar{R}){{\cal U}}$. 
Then the variation of $\bar{\Phi}$ is $\delta \bar{\Phi}= 
({{\cal D}}{{\cal D}}- {R})\delta {{\cal U}}$].}
\begin{eqnarray}\label{varEa}
\delta E^{a} & = E^a (\Lambda (\delta ) + \bar{\Lambda} (\delta )) 
 - {1\over 4} E^b \tilde{\sigma}_b^{ \dot{\alpha} {\alpha} }
[{\cal D}_{{\alpha}}, \bar{{\cal D}}_{\dot{\alpha}}] \delta H^a + 
 \hspace{-0.5cm} 
\nonumber \\ 
& + i E^{\alpha} {\cal D}_{{\alpha}}\delta H^a   
- i \bar{E}^{\dot{\alpha}}\bar{{\cal D}}_{\dot{\alpha}} \delta H^a \; , 
\\ \label{varEal}
 \delta E^{\alpha} & =  E^a \Xi_a^{\alpha}(\delta ) + 
E^{\alpha} \Lambda (\delta ) 
+ {1\over 8} \bar{E}^{\dot{\alpha}} R \sigma_a{}_{\dot{\alpha}}{}^{\alpha}
\delta H^a \; .  
\end{eqnarray} 
In Eqs. (\ref{varEa}), (\ref{varEal}),  
$\Lambda (\delta )$, $\bar{\Lambda} (\delta )$ are given by 
\begin{eqnarray}\label{Lb}
\Lambda (\delta ) & = {1\over 24} 
\tilde{\sigma}_a^{ \dot{\alpha} {\alpha} }
[{\cal D}_{{\alpha}}, \bar{{\cal D}}_{\dot{\alpha}}] \delta H^a + 
{i\over 4}{\cal D}_a  \delta H^a + {1\over 24} G_a   \delta H^a 
\nonumber \\ & + 2 ( {\cal D}{\cal D}- \bar{R})\delta {\cal U} 
- ( \bar{{\cal D}}\bar{{\cal D}}- {R})\delta \bar{{\cal U}} \; 
\\ \label{Lb+cc}
\Lambda (\delta ) & + \bar{\Lambda} (\delta )  =
{1\over 12} 
\tilde{\sigma}_a^{ \dot{\alpha} {\alpha} }
[{\cal D}_{{\alpha}}, \bar{{\cal D}}_{\dot{\alpha}}] \delta H^a + 
{1\over 12} G_a   \delta H^a + \nonumber \\ 
& + ( {\cal D}{\cal D}- \bar{R})\delta {\cal U} 
+ ( \bar{{\cal D}}\bar{{\cal D}}- {R})\delta \bar{{\cal U}} \; ;  
\end{eqnarray}
the explicit expression for $\Xi_a^{\alpha}(\delta )$ 
in (\ref{varEal}) will not be needed below. It reads
\begin{eqnarray}\label{Xi3/2}
\Xi_a^{\alpha}(\delta )& = {i\over 4} {\sigma}_{a\beta \dot{\gamma}}
u^{\dot{\gamma}\; \alpha\beta} (\delta)  
- {i\over 4} \tilde{\sigma}_a^{ \dot{\alpha} {\alpha} } 
\bar{{\cal D}}_{\dot{\alpha}} \Lambda (\delta ) - 
\qquad \nonumber \\ & - 
{i\over 32} {\sigma}_{a \beta \dot{\beta}} 
{\cal D}^{\beta} R \delta H^{\alpha \dot{\beta}} - 
{i\over 16} {\sigma}_{a \beta \dot{\beta}} 
R {\cal D}^{\beta}\delta H^{\alpha \dot{\beta}} 
- \nonumber \\ & \quad 
- {i\over 32} \tilde{\sigma}_a^{\dot{\beta}\beta} G^{\alpha\dot{\gamma}}
 \bar{{\cal D}}_{\dot{\beta}} \delta H_{\beta \dot{\gamma}} \; , 
\end{eqnarray}
where 
\begin{eqnarray}\label{uadd}
& u_{\dot{\gamma}}^{\alpha\beta}(\delta)  =  -{1\over 4} 
\bar{{\cal D}}\bar{{\cal D}} {\cal D}^{(\alpha} 
\delta H^{\beta)}{}_{\dot{\gamma}} + {3\over 8} R  {\cal D}^{(\alpha} 
\delta H^{\beta)}{}_{\dot{\gamma}} 
-  \qquad \nonumber \\ & \qquad 
-{1\over 8} G^{(\alpha}{}_{\dot{\beta}} \bar{{\cal D}}_{\dot{\gamma}}
\delta H^{\beta)\dot{\beta}} 
+ {1\over 16} {\cal D}^{(\alpha} R \, \delta H^{\beta)}{}_{\dot{\gamma}} 
-  \qquad \nonumber \\ & \qquad 
- {1\over 8}\bar{{\cal D}}_{(\dot{\gamma}} G^{(\alpha}{}_{\dot{\beta})}\, 
\delta H^{\beta)\dot{\beta}} + {1\over 8} W^{\alpha\beta\gamma} 
\delta H_{\gamma\dot{\gamma}} \; . \qquad  
\end{eqnarray}
Eq. (\ref{uadd}), together with 
\begin{eqnarray}\label{uabc}
& u_{{\gamma}}^{\alpha\beta}(\delta)  =  
{1\over 8} G^{(\alpha}{}_{\dot{\beta}} {\cal D}_{\gamma}
\delta H^{\beta)\dot{\beta}} - 
{1\over 8} G_{\delta\dot{\delta}} \delta_{\gamma}^{(\alpha} {\cal D}^{\beta)}
\delta H^{\delta\dot{\delta}} + \nonumber \\ & \qquad {} \qquad + 
2\delta_{\gamma}^{(\alpha} {\cal D}^{\beta)} \Lambda (\delta) \; ,  
\\ & \sigma^a_{\gamma\dot{\gamma}}
u_{a}^{\alpha\beta}(\delta)   = - {i\over 2}(
{\cal D}_{\gamma}u_{\dot{\gamma}}^{\alpha\beta}(\delta) + 
\bar{{\cal D}}_{\dot{\gamma}}u_{{\gamma}}^{\alpha\beta}(\delta)) - 
  \qquad {} \nonumber \\ & \qquad
-{i\over 16} R{\bar R}  \delta_{\gamma}^{(\alpha} \delta H^{\beta)}{}_{\dot{\gamma}}
-  {i\over 16} \bar{{\cal D}}_{\dot{\beta}}\bar{R}  
\delta_{\gamma}^{(\alpha} \bar{{\cal D}}_{\dot{\gamma}}  
\delta H^{\beta)\dot{\beta}} +  \nonumber \\ & \qquad +
{i\over 16}{\cal D}^{(\alpha}G^{\beta)\dot{\beta}} 
\bar{{\cal D}}_{\dot{\gamma}} \delta H_{\gamma\dot{\beta}} + 
{i\over 16} W^{\alpha\beta\delta} {\cal D}_{\gamma} 
\delta H_{\delta\dot{\gamma}} \; , 
\end{eqnarray}
define the variation  of the spin connection 
through the second equation in (\ref{varEMA}).

\subsection{Superfield action and `free' equations  of motion}

The nontrivial dynamical equations of motion should follow from the 
variations (\ref{varEa}), (\ref{varEal}) with (\ref{Lb}), (\ref{Lb+cc})
only. 
The variation of the superdeterminant $E=sdet(E_M^A)$ 
under  (\ref{varEa}), (\ref{varEal}), has the form 
(see \cite{WZ78})  
\begin{eqnarray}\label{varsdE}
\delta E = & E [ - {1\over 12} \tilde{\sigma}_a^{\dot{\alpha}\alpha} [
{\cal D}_{\alpha}, \bar{\cal D}_{\dot{\alpha}}] \delta H^a + 
{1\over 6} G_a \; \delta H^a + 
 \nonumber \\  
& + 2(\bar{\cal D} \bar{\cal D} - R) \delta \bar{{\cal U}} +  
2({\cal D} {\cal D} - \bar{R}) \delta {\cal U}] \; .
\end{eqnarray}
In the light of the identity (\ref{idd8}),  
all the terms with derivatives can be omitted in (\ref{varsdE})
 when one considers the variation of the action (\ref{SGact}). 
Hence,
\begin{eqnarray}\label{vSGsf}
& \delta S_{SG} = \int d^8 Z \; \delta E = \nonumber \\ & = 
\int d^8Z E\;  [{1\over 6} G_a \; \delta H^a -  
2 R\; \delta \bar{{\cal U}}
-2 \bar{R}\; \delta {\cal U}] \; 
\end{eqnarray}
and one arrives at the following {\sl superfield equations of 
motion for `free', simple $D=4$ $N=1$ supergravity}:  
\begin{eqnarray}\label{SGeqmG}
 {\delta S_{SG}\over \delta H^a}=0 \quad \Rightarrow \quad  G_a =0 \; ,  
\\ \label{SGeqmR}
{\delta S_{SG}\over \delta \bar{{\cal U}}}=0 \quad \Rightarrow \quad
 R=0 \; , 
\\ \label{SGeqmbR} 
{\delta S_{SG}\over \delta {{\cal U}}}=0 \quad \Rightarrow 
\quad \bar{R}=0 \; .  
\end{eqnarray}
Then the {\sl `free' superfield Rarita--Schwinger equations}, 
\begin{eqnarray}\label{SGRS}
& \epsilon^{abcd} T_{bc}{}^\gamma \sigma_{d\gamma\dot{\gamma}} =0 \; , 
\qquad 
\epsilon^{abcd} T_{bc}{}^{\dot{\gamma}} \sigma_{d\gamma\dot{\gamma}} =0 \; ,
\end{eqnarray}
follow from the constraints (\ref{Tabg}), (\ref{Tabdg})
with $G_a=0=R$,  
\begin{eqnarray}\label{TabgW}
& T_{{\alpha}\dot{\alpha}\; \beta \dot{\beta }\; {\gamma}} 
= - {1\over 8} \epsilon_{\dot{\alpha}\dot{\beta}}W_{\alpha \beta\gamma}\; 
\qquad \nonumber \\ & \Leftrightarrow \qquad 
T_{ab}{}^\gamma = {1\over 32}(\sigma_a \tilde{\sigma}_b)_{\alpha\beta}
W^{\alpha\beta\gamma} \; . 
\end{eqnarray}
The superfield generalization of the free Einstein equation 
$R_{ac}{}^{bc}= {1\over 2} \delta_a^b R_{cd}{}^{cd}=0$ follows from 
setting $G_a=0$ (Eq. (\ref{SGeqmG})) 
and $R=0$ (Eq.  (\ref{SGRS})) in Eq. (\ref{RRici}).

\section{BRINK--SCHWARZ SUPERPARTICLE IN A SUPERGRAVITY BACKGROUND}

\renewcommand{\theequation}{\arabic{section}.\arabic{equation}} 
\setcounter{equation}0 

The superparticle dynamical variables are the 
supercoordinate functions $\hat{Z}^{ {M}}(\tau )$ defined 
by the map  
\begin{eqnarray}\label{W1s}
\hat{\phi}: W^1\rightarrow {\Sigma}^{(4|4)}\; , 
\;\; \tau \mapsto \hat{Z}^{ {M}}(\tau )=   
(\hat{x}^{ {\mu}}(\tau), \, \hat{\theta}^{\breve{\alpha}}(\tau) 
\,) \; . 
\end{eqnarray} 
defining a worldline ${\cal W}^1$ in ${\Sigma}^{(4|4)}$     
parametrized by the proper time  
$\tau$, 
\begin{eqnarray}\label{M1}
{\cal W}^1 \subset {\Sigma}^{(4|4)} \; , & \quad 
Z^M= \hat{Z}^M(\tau) \; . 
\end{eqnarray}
The actual 
superparticle worldline 
is determined by the equations of motion. 
For the massless superparticle  these equations 
follow from 
the Brink--Schwarz action 
\begin{equation}\label{BSac}
S_{sp} = \int_{W^1} \hat{{\cal L}}_1 = 
{1 \over 2} \int_{W^1} l(\tau ) \hat{E}^a 
  \hat{E}^b_{\tau}\eta_{ab}\; , \qquad  
\end{equation}
which involves the pull--back $\hat{E}^a\equiv \hat{E}^a(\tau)= 
d{\tau}\hat{E}_{\tau}^a(\tau)$ 
to $W^1$ 
of the bosonic supervielbein form $E^a$ (Eq. (\ref{4Ea})) on 
${\Sigma}^{(4|4)}$,  
\begin{eqnarray}\label{hatEa}
  \hat{E}^{ {a}} = d\hat{Z}^{M}(\tau) E_{M}^{~ {a}}(\hat{Z})
\equiv  d\tau \hat{E}^{ {a}}_\tau 
\\ 
\nonumber 
\hat{E}^{ {a}}_\tau= 
\partial_\tau \hat{Z}^{M} E_{M}^{~ {a}}(\hat{Z})
\;     
\end{eqnarray}  
and the Lagrange multiplier (worldline einbein) $l(\tau )$. 
Note that the pull--backs of the fermionic supervielbein forms 
\begin{eqnarray}\label{hatEal}
  & \hat{E}^{ {\alpha}} = d\hat{Z}^{M}(\tau) E_{M}^{~ {\alpha}}(\hat{Z})
\equiv  d\tau \hat{E}^{ {\alpha}}_\tau \; , \qquad 
\nonumber \\ & 
\hat{\bar{E}}{}^{ \dot{\alpha}} = d\hat{Z}^{M}(\tau) 
\bar{E}_{M}^{~ \dot{\alpha}}(\hat{Z})
\equiv  d\tau \hat{\bar{E}}{}^{ \dot{\alpha}}_\tau \; , \qquad 
\\ \nonumber  & 
\hat{E}^{  {\alpha}}_\tau= 
\partial_\tau \hat{Z}^{M} E_{M}^{~{\alpha}}(\hat{Z}) \; , 
    \qquad  \hat{\bar{E}}{}^{\dot{\alpha}}_\tau= 
\partial_\tau \hat{Z}^{M} \bar{E}_{M}^{~\dot{\alpha}}(\hat{Z})
\;      
\end{eqnarray}  
are not involved in the superparticle action (\ref{BSac}) explicitly. 
This is a general property of the $D$--dimensional super--$p$--brane actions 
that reflects an especial r\^ole for the bosonic `directions' 
in superspace.

\subsection{Equations of motion}

The equations of motion for a superparticle moving in a supergravity 
{\sl background}  
follow from the variation of the action (\ref{BSac}) with respect to 
the Lagrange multiplier, $\delta l(\tau )$ and the supercoordinate functions,  
$\delta \hat{Z}$. The corresponding 
variation of the pull--back (\ref{hatEa}) of the bosonic supervielbein
form (\ref{4Ea}) is 
 \begin{eqnarray}
\label{*gcEa} 
\delta_{\hat{Z}} \hat{E}^{a} & \equiv 
\delta_{\hat{Z}} {E}^{a}(\hat{Z}) :=  E^{a} (\hat{Z}+ \delta\hat{Z}) - 
E^{a}(\hat{Z})= 
\nonumber \\ & =  i_{\delta \hat{Z}} \hat{T}^{a} + 
{\cal D}(i_{\delta \hat{Z}} \hat{E}^a) + 
\hat{E}^{b} i_{\delta \hat{Z}} w_{b}{}^{a} \; , 
\\ \label{iZEa} 
& i_{\delta \hat{Z}} E^{a}(\hat{Z}):= \delta \hat{Z}^M E_M^{a}(\hat{Z}) \; , 
\qquad  \\ \label{iZwab} 
& i_{\delta \hat{Z}} w^{ab}:=  \delta \hat{Z}^M\; w_M^{ab}(\hat{Z})\; 
\end{eqnarray}
(note in passing that these transformations 
coincide with the pull--back of superspace 
general coordinate transformations $\delta_{gc}$, 
Eqs. (\ref{gcZ}), (\ref{gcEA}), 
to $W^1$, $\delta_{\hat{Z}} \hat{E}^a = 
\hat{\phi}^{*}(\delta_{gc}E^a(Z))$). 

The last term in (\ref{*gcEa}) does not contribute to the action 
variation 
\footnote{ 
This reflects the invariance of the action under a Lorentz rotation 
of the supervielbein, which can be considered as a pull--back of 
the local Lorentz transformation of the supergravity background. 
Such transformations cannot be treated as {\sl gauge} symmetries 
of the superparticle in a supergravity {\sl background}. 
However, they {\sl are}  gauge symmetries of the interacting system 
of {\sl dynamical} supergravity and the superparticle.}
\begin{eqnarray}\label{vBSac0}
\delta S_{sp}  
 & = \int_{W^1} [{1 \over 2} 
\delta l(\tau ) & \hat{E}_{\tau a} 
  \hat{E}^a \; +  l(\tau ) \; \hat{E}_{\tau a} 
 \delta_{\hat{Z}} \hat{E}^a \;] \; , \qquad  
\end{eqnarray}
 because $ \hat{E}_{\tau a}  \hat{E}_{\tau b} i_{\delta \hat{Z}} w^{ba}
\equiv 0$ due to $i_{\delta \hat{Z}} w^{ba}= 
- i_{\delta \hat{Z}} w^{ab}$.

When the background obeys the constraints 
(\ref{4WTa=}) the superparticle equations of motion become
\begin{eqnarray}\label{fEqm}
&
\hat{E}^\alpha  \sigma^a_{\alpha\dot{\alpha}} \hat{E}_{a\tau} =0 \; , \qquad 
\hat{E}_{a\tau} \sigma^a_{\alpha\dot{\alpha}} \hat{\bar{E}}{}^{\dot{\alpha}} 
= 0 \; ,  \\ 
\label{bEqm}
& {\cal D}(l\, \hat{E}_{a\tau}) = 0\; , \\ 
\label{lEqm}
& \hat{E}^a_{\tau} \hat{E}_{a\tau}=0\; .
\end{eqnarray}

Indeed, Eq. (\ref{4WTa=}) implies 
\begin{eqnarray}\label{iZTa}
i_{\delta\hat{Z}}\hat{T}^a & = 
- 2i\sigma^a_{\alpha\dot{\alpha}} 
\hat{E}^\alpha i_{\delta\hat{Z}} \hat{\bar{E}}{}^{\dot{\alpha}} 
- 2i\sigma^a_{\alpha\dot{\alpha}} 
\hat{\bar{E}}{}^{\dot{\alpha}}  i_{\delta\hat{Z}} \hat{E}^\alpha -
\nonumber \\ & - {1\over 8} \hat{E}^b  \varepsilon^a{}_{bcd} 
G^c(\hat{Z}) \; i_{\delta\hat{Z}} \hat{E}^d \; .  
\end{eqnarray}
The last term does not contribute to the 
contraction $\hat{E}_{\tau a} i_{\delta\hat{Z}}\hat{T}^a$. Hence, 
after integration by parts, the expression  (\ref{vBSac0}) with 
(\ref{*gcEa}) becomes 
\begin{eqnarray}\label{vBSac1}
\delta S_{sp}  & = \int_{W^1} [ 
{1 \over 2} 
\delta l\,  \hat{E}_{\tau a} 
  \hat{E}^a \; 
- {\cal D} (l\,  \hat{E}_{\tau a}) \; 
i_{\delta \hat{Z}} \hat{E}^a - \nonumber \\  
& - 2i \, l \, \hat{E}_{\tau a} 
(\sigma^a_{\alpha\dot{\alpha}} 
\hat{E}^\alpha i_{\delta\hat{Z}} \hat{\bar{E}}{}^{\dot{\alpha}} 
+  \sigma^a_{\alpha\dot{\alpha}} 
\hat{\bar{E}}{}^{\dot{\alpha}}  i_{\delta\hat{Z}} \hat{E}^\alpha ) ]\; , 
\end{eqnarray}
which implies the equations of motion 
(\ref{lEqm}) (${\delta S_{sp}\over \delta l}=0$), 
(\ref{bEqm}) (${\delta S_{sp}\over \delta \hat{Z}^M} E_a^M(\hat{Z})=0$)
and  (\ref{fEqm}) 
(${\delta S_{sp}\over \delta \hat{Z}^M} E_{\dot{\alpha}}^M(\hat{Z})=0$ 
and its complex conjugate).

Let us stress that we derived the superparticle equations 
of motion  (\ref{bEqm}), (\ref{fEqm}) from an arbitrary variation of 
the supercoordinate functions $\delta\hat{Z}$, which is tantamount to saying 
that they were obtained 
{\sl from the general coordinate transformations $\delta_{gc}$, 
 (\ref{gcZ}), (\ref{gcEA}), pulled--back  to $W^1$}. 
This reflects a {\sl spontaneous} (partial) breaking of the 
superspace general coordinate symmetry $\delta_{gc}$ 
of the background by the superparticle worldline. The 
part of the general coordinate symmetry 
$\delta_{gc}$ of the supergravity background 
which is preserved by the worldline, can be identified with 
the gauge fermionic $\kappa$--symmetry \cite{ALS} and reparametrization 
symmetry (more rigorously, the variational version of the 
worldline general coordinate symmetry) \cite{BdAI2}.

\subsection{Local fermionic $\kappa$--symmetry and reparametrization 
invariance of superparticle action}

It is not hard to see that the  superparticle action (\ref{BSac})
is invariant under the gauge fermionic  $\kappa$--symmetry \cite{ALS} 
that acts on the coordinate functions and the Lagrange multiplier $l$ by 
\begin{eqnarray}\label{4kappa}
\delta_\kappa \hat{Z}^M =
\hat{ E}^a_{\tau} \tilde{\sigma}_a^{\dot{\alpha} {\alpha}} 
(\bar{\kappa}_{\dot{\alpha}}(\tau) \, E_\alpha^M(\hat{Z}) 
+ {\kappa}_{{\alpha}}(\tau) \, \bar{E}_{\dot{\alpha}}^M(\hat{Z})) 
\; ,  
 \\ 
\label{kapl}
 \delta_{\kappa}l(\tau)=4i l\, 
(\hat{E}_\tau^{{\alpha}}
\, {\kappa}_{{\alpha}}(\tau) + 
\hat{\bar{E}}{}^{\dot{\alpha}}_\tau
\, \bar{\kappa}_{\dot{\alpha}}(\tau))
\; ,  \qquad 
\end{eqnarray}
To this end, it is convenient to write (\ref{4kappa}) in the form 
\begin{eqnarray}\label{kappa}
& i_\kappa \hat{E}^a \equiv \delta_\kappa \hat{Z}^M  E_M^{\; a}(\hat{Z})=0\; , 
\\ \label{kapf}  
& i_\kappa \hat{E}^{{\alpha}}\equiv
\delta_\kappa \hat{Z}^M  E_M^{\; {\alpha}}(\hat{Z})=
\bar{\kappa}_{\dot{\alpha}}(\tau)\tilde{\sigma}_a^{ \dot{\alpha} {\alpha}} 
\hat{ E}^a_{\tau} \; , \nonumber \\  
& i_\kappa \hat{\bar{E}}{}^{\dot{\alpha}}\equiv
\delta_\kappa \hat{Z}^M  E_M^{\; \dot{\alpha}}(\hat{Z})=
\hat{ E}^a_{\tau} \tilde{\sigma}_a^{\dot{\alpha} {\alpha}} 
{\kappa}_{{\alpha}}(\tau)\; ,
\qquad 
\end{eqnarray}
substitute these $i_\kappa \hat{E}^{A}$ and $\delta_{\kappa}l(\tau)$ 
(Eq. (\ref{kapl})) for 
 $i_{\delta \hat{Z}} \hat{E}^{A}$ and $\delta l(\tau)$ in (\ref{vBSac1}), 
and observe that, due to the identity  
\begin{eqnarray}\label{idEE}
\hat{E}_{a\tau} \hat{E}_{b\tau} \; 
(\sigma^a\tilde{\sigma}^b)_{{\alpha}}{}^{{\beta}} 
= \hat{E}^a_{\tau} \hat{E}_{a\tau} 
\delta_{{\alpha}}{}^{\beta} \; ,  
\end{eqnarray}
the contribution  
$l \hat{E}_{a\tau} \delta_\kappa \hat{E}^a  
= -2i l \hat{E}_{a\tau} \hat{E}^a_{\tau} 
\, \hat{E}^{\underline{\alpha}}  \kappa_{\underline{\alpha}}
+ c.c.\;$ 
can be compensated by the variation of the Lagrange multiplier 
$\delta_{\kappa}l$ (\ref{kapl}).

In the same manner, one finds that the 
following transformations of the supercoordinate functions 
\begin{eqnarray}\label{4rep}
\delta_r \hat{Z}^M = r(\tau) \hat{E}^a_{\tau} E_a^M (\hat{Z}) \; , 
\end{eqnarray}
or, equivalently, 
\begin{eqnarray}\label{rep}
& i_r \hat{E}^a \equiv \delta_r \hat{Z}^M  E_M^{\; a}(\hat{Z})=
r(\tau) \hat{ E}^a_{\tau}
\; , \quad \nonumber \\ &  i_r \hat{E}^{{\alpha}}=0\; , \qquad 
i_r \hat{\bar{E}}{}^{\dot{\alpha}}=0 \; ,
\end{eqnarray}
can be compensated by
\footnote{On the worldvolume, acting on the pull--back 
of the superforms,  
${\cal D}= d\hat{Z}^M {\cal D}_M  
=d\tau {\cal D}_\tau$, where ${\cal D}_\tau = \partial_\tau + connection 
\; term(s)$.
 Integrating by parts 
one arrives at the terms 
involving $\partial_\tau l$ in the worldvolume action variations. 
Note that ${\cal D}_\tau l = \partial_\tau l$, because the 
einbein $l(\tau)$ does not have Lorentz group indices.}
\begin{eqnarray}\label{repl}
 \delta_{r}l(\tau)= l \partial_\tau r - r \partial_\tau l \; .
\end{eqnarray}
This proves the so--called reparametrization symmetry of the superparticle 
action (see also Appendix D).

\section{Complete Lagrangian description of the 
SUPERGRAVITY--SUPERPARTICLE INTERACTING SYSTEM} 

\renewcommand{\theequation}{\arabic{section}.\arabic{equation}} 
\setcounter{equation}0

A fully dynamical description of $D=4$ $N=1$ 
supergravity and the massless superparticle source 
interacting system 
can be achieved by means of the action 
\begin{eqnarray}\label{SGSP}
S=S_{SG} + S_{sp}  
= \int d^8 z \; E
+ {1 \over 2} \int_{W^1} l(\tau ) \hat{E}^a   \hat{E}^b_{\tau}\eta_{ab}\; ,  
\end{eqnarray} 
where $E= sdet(E_M^A)$ and the supervielbein in superspace is assumed to 
be restricted by the constraints (\ref{4WTa=}), (\ref{4WTal=}), 
(\ref{4WTdA=}).

\subsection{Gauge symmetries of the coupled system}

As the superparticle coordinate functions
$\hat{Z}^M\equiv \hat{Z}^M(\tau)$ do not enter in the supergravity 
part of the action, Eqs. (\ref{fEqm}), (\ref{bEqm}), (\ref{lEqm})
remain the same as in the interacting system (\ref{SGSP}), 
\begin{eqnarray}\label{fEqmi}
&
\hat{E}^\alpha  \sigma^a_{\alpha\dot{\alpha}} \hat{E}_{a\tau} =0 \; , \qquad 
\\\label{fbEqmi} & \hat{E}_{a\tau} \sigma^a_{\alpha\dot{\alpha}} 
\hat{\bar{E}}{}^{\dot{\alpha}} = 0 \; , \qquad  \\ 
\label{bEqmi}
& {\cal D}(l\, \hat{E}_{a\tau}) = 0\; ,   
\\ 
\label{lEqmi}
& \hat{E}^a_{\tau} \hat{E}_{a\tau}=0\; .
\end{eqnarray} 
Moreover, $\kappa$--symmetry (Eqs. (\ref{kappa}), (\ref{kapf}), 
(\ref{kapl})) and  reparametrization symmetry (Eqs. 
(\ref{rep}), (\ref{repl})) are preserved by the interaction.

The coupled action is evidently invariant under superdiffeomorphisms 
$\delta_{diff}$, 
\begin{eqnarray}
\label{sdZ}
& Z^{\prime M}= Z^M + b^M(Z)\; : \quad 
\cases{
x^{\prime \mu}= x^\mu + b^\mu (x, \theta )\, , \cr 
\theta^{\prime \breve{\alpha}}=  \theta^{\breve{\alpha}} +  
\varepsilon^{\breve{\alpha}}
(x, \theta) \; , }\, 
\\ \label{sdiffF}
& E^{\prime A}(Z^\prime) =  
E^{A}(Z), \quad  w^{\prime ab}(Z^\prime)=
w^{ab}(Z)\; ,   \qquad 
\end{eqnarray} 
 now supplemented by the corresponding transformations for the 
superparticle variables   $\hat{Z}^{\prime M}=\hat{Z}^{\prime M}(\tau)$ 
\begin{eqnarray}
\label{sdhZ} 
& \hat{Z}^{\prime M}= \hat{Z}^M + b^M(\hat{Z})\; : \; \cases{
\hat{x}^{\prime \mu}(\tau)= 
\hat{x}^\mu + b^\mu (\hat{x} , \hat{\theta})\, , \cr 
 \hat{\theta}^{\prime \breve{\alpha}}(\tau) =  
\hat{\theta}^{\breve{\alpha}} +  
\varepsilon^{\breve{\alpha}}
(\hat{x}, \hat{\theta})\;  ,  } 
\end{eqnarray} 
so that 
\begin{eqnarray}
\label{sdiffZ}
& \delta_{diff}Z^M=Z^{\prime M} - Z^M= b^M(Z)\; , \; 
\\ 
\label{sdiffhZ}
& \delta_{diff} \hat{Z}^{M}=  b^M(\hat{Z}) \; 
\end{eqnarray} 
and $\delta_{diff}S=0$ (see Appendix B2, where further  
discussion on the gauge symmetries of the coupled system 
can be found).

\subsection{Equations of motion of the coupled system}

As mentioned above, the superparticle equations 
$\delta S/\delta \hat{Z}^M=0,\,\delta S/\delta l=0$ for 
the coupled dynamical system remain the same 
as those for the system in a supergravity {\sl background},  
$\delta S_{sp}/\delta \hat{Z}^M=0,\,\delta S_{sp}/\delta l=0$  
(Eqs. (\ref{fEqmi}), (\ref{fbEqmi}), (\ref{bEqmi}) and
(\ref{lEqmi})). Let us now see how the supergravity 
equations of motion are modified by the inclusion of the 
superparticle source.

Denoting the variation of the action 
induced by the constraints preserving variations 
(\ref{varEa})--(\ref{Lb+cc}) by $\delta^\prime$ ,  
one concludes that 
\begin{eqnarray}\label{vSsf}
\delta^\prime S  & =  
 \int d^8Z E\;  [{1\over 6} G_a \; \delta H^a -  
2 R\; \delta \bar{{\cal U}}
- 2 \bar{R}\; \delta {\cal U}] + \nonumber 
\\ & \qquad +  \delta^\prime S_{sp} \; ,
\end{eqnarray}
where 
\begin{eqnarray}\label{vSsf1}
\delta^\prime S_{sp} & = \int_{W^1}  
 l(\tau ) \hat{E}_{a\tau} \delta^\prime \hat{E}^a
= \qquad \nonumber \\ &  = \int_{W^1}  
 l(\tau ) \hat{E}_{a\tau} d\hat{Z}^M \delta^\prime E_M^a(\hat{Z})
\end{eqnarray}
and  $\delta^\prime \hat{E}^a$ is the pull--back of 
(\ref{varEa}) to $W^1$. To have a 
well posed variational problem, 
we extend the integration in (\ref{vSsf1}) to 
superspace by introducing the superspace delta--function 
\begin{eqnarray}\label{delta}
& \delta^8(Z-\hat{Z}) := \delta^4(x-\hat{x}) 
(\theta-\hat{\theta})^4 
\end{eqnarray}
where
\begin{eqnarray}\label{delta8}
& (\theta-\hat{\theta})^4 := 
{1\over 4!} \epsilon_{\breve{\alpha}_1\ldots \breve{\alpha}_4}
 (\theta-\hat{\theta})^{\breve{\alpha}_1}\ldots 
(\theta-\hat{\theta})^{\breve{\alpha}_4}\; .
\end{eqnarray}
Namely, we insert  
$1= \int d^8 Z  \delta^8(Z-\hat{Z})$ into 
(\ref{vSsf1}) and use the identity 
$\delta^\prime E_M^a(\hat{Z}) \delta^8(Z-\hat{Z})
\equiv \delta^\prime E_M^a(Z) \delta^8(Z-\hat{Z})$
to arrive at 
\begin{eqnarray}\label{vSsf2}
& \delta^\prime S_{sp}= 
\int d^8 Z \, [\int_{W^1}  
 l(\tau ) \hat{E}_{a\tau} d\hat{Z}^M \delta^8(Z-\hat{Z}) ]
\, \delta^\prime E_M^a(Z) \; . \hspace{-0.9cm}  \nonumber \\ 
\end{eqnarray}
Now Eq. (\ref{varEa}) can be straightforwardly inserted into 
(\ref{vSsf2}) and, using 
\begin{eqnarray}\label{id81}
& d\hat{Z}^M \delta^8(Z-\hat{Z}) E_M^A(Z) 
 \equiv \hat{E}^A \delta^8(Z-\hat{Z}) \equiv \nonumber \\ 
& \qquad {}\equiv E(Z)  {1\over \hat{E}}   \hat{E}^A \delta^8(Z-\hat{Z})\; 
\; , \\ \nonumber
& E(Z)\equiv sdet(E_M^A(Z))\; ,  \quad  \hat{E}= E(\hat{Z})\; , 
\end{eqnarray}
one finds 
\begin{eqnarray}
\label{vSsf3}
& \delta^\prime S_{sp}= \qquad {}\qquad {}\qquad {}\qquad {}\qquad {}\qquad {}
\nonumber \\ &
= \int d^8 Z E  [\int_{W^1}  
 {l(\tau )\over \hat{E}} \hat{E}_{a\tau} \hat{E}^{\alpha} 
\delta^8(Z-\hat{Z}) ]
i {\cal D}_{\alpha} \delta H^a \; + \hspace{-0.9cm}  \nonumber \\  
&
+ \int d^8 Z E   [ \int_{W^1}  
 {l(\tau )\over \hat{E}} \hat{E}_{a\tau} \hat{E}^{\dot{\alpha}}
\delta^8(Z-\hat{Z}) ]
(- i) \bar{{\cal D}}_{\dot{\alpha}} \delta H^a \; - \hspace{-0.9cm}  
\nonumber \\ 
&
- \int d^8 Z E   [ \int_{W^1}  
 {l(\tau )\over \hat{E}} \hat{E}_{a\tau} \hat{E}^{b}
\delta^8(Z-\hat{Z}) ] \times \qquad \nonumber \\ 
& \qquad {} \qquad \times 
 {1\over 4} \tilde{\sigma}_b^{ \dot{\alpha} {\alpha} }
[{\cal D}_{{\alpha}}, \bar{{\cal D}}_{\dot{\alpha}}] \delta H^a 
+ \hspace{-0.9cm}  \nonumber \\
& + \int d^8 Z E   [ \int_{W^1}  
 {l(\tau )\over \hat{E}} \hat{E}_{a\tau} \hat{E}^{a}
\delta^8(Z-\hat{Z}) ] \, 
(\Lambda (\delta ) + \bar{\Lambda} (\delta ))
\; .   \hspace{-0.9cm}   \nonumber \\ 
\end{eqnarray}
The extraction of the superdeterminant 
in (\ref{vSsf3}) permits integrating  
 by parts using the identity (\ref{idd8}) in 
Appendix B. Thus Eqs. (\ref{vSsf}), (\ref{vSsf3}) allows us 
a direct derivation of the coupled equations of motion. 

Note that the scalar variations $\delta {\cal U}$, 
$\delta \bar{{\cal U}}$ are involved only in the last term of 
(\ref{vSsf}), through $(\Lambda (\delta ) + \bar{\Lambda} (\delta ))$ 
defined by (\ref{Lb+cc}).

Let us now compute the $\delta {\cal U}$ variation of the coupled action,  
$\delta_{{\cal U}}S = \delta_{{\cal U}}S_{SG}+ \delta_{{\cal U}}S_{sp}$. 
The variation of the supergravity part reads 
$ \delta _{{\cal U}} S_{SG} =  -  {1\over 2}   \int d^8Z \; E \; 
\bar{R}\; \delta {\cal U}\;$ (see Eq. (\ref{vSGsf})), while, due to 
$(\Lambda (\delta_{{\cal U}} ) + \bar{\Lambda} (\delta_{{\cal U}} ))= 
({\cal D}{\cal D}-\bar{R})\delta {\cal U}\;$ (see \ref{Lb+cc}),  
\begin{eqnarray}\label{vSspU}
& \delta_{{\cal U}} S_{sp} = 
\nonumber \\ & = \int d^8 Z E   [ \int_{W^1}
 {l \over \hat{E}} \hat{E}_{a\tau} \hat{E}^{a}
\delta^8(Z-\hat{Z}) ]  ({\cal D}{\cal D}-\bar{R})\delta {\cal U}
\nonumber \\ & = 
 \int d^8 Z E   [ \int_{W^1}  
 {l\over \hat{E}} \hat{E}_{a\tau} \hat{E}^{a}
({\cal D}{\cal D}-\bar{R}) \delta^8(Z-\hat{Z}) ]  
\delta {\cal U}\;  . 
\end{eqnarray}
Thus, at a first look, Eq. (\ref{SGeqmbR}) acquires a source term 
\begin{eqnarray}\label{SGeqmbR+0} 
& {\delta S \over \delta {{\cal U}}}=0 \quad \Rightarrow 
\quad \bar{R}= {\cal J}_0 \; \nonumber \\ 
& {\cal J}_0 =  
\int_{W^1}  
 {l\over 2 \hat{E}} \hat{E}_{a\tau} \hat{E}^{a}
({\cal D}{\cal D}-R) \delta^8(Z-\hat{Z})   \; . \hspace{-0.9cm}  
\nonumber \\ 
\end{eqnarray}
However, one immediately observes that this source vanishes 
due the superparticle equation of motion (\ref{lEqmi})
\begin{eqnarray} 
\label{J0=0}
& {\delta S \over \delta l(\tau)}= 0 \quad \Rightarrow \quad 
 \hat{E}^a_{\tau} \hat{E}_{a\tau}=0 \quad  \Rightarrow \quad 
{\cal J}_0 =0 \; .
\end{eqnarray}
Hence, {\sl the  scalar superfield equations for the coupled 
dynamical system are the same as in the `free' supergravity case}, 
\begin{eqnarray} \label{SeqmbR} 
{\delta S \over \delta {{\cal U}}}=0 \quad \Rightarrow 
\quad \bar{R}=0 \; ,   
\\ \label{SeqmR}
{\delta S \over \delta \bar{{\cal U}}}=0 \quad \Rightarrow \quad
 R=0 \; .
\end{eqnarray}
 
Moreover, the above observation implies that the last term 
in the superparticle action variation does not contribute to the 
equations of motion, 
\begin{eqnarray}\label{intid} 
& \int d^8 Z E   [ \int_{W^1}  
 {l(\tau )\over \hat{E}} \hat{E}_{a\tau} \hat{E}^{a}
\delta^8(Z-\hat{Z}) ] \, 
(\Lambda (\delta ) + \bar{\Lambda} (\delta ))= 0\, ,
\hspace{-0.9cm} \nonumber \\ & 
\end{eqnarray}
due to Eq.  (\ref{lEqmi}), $\hat{E}^a \hat{E}_{a\tau}=0$. 
Hence, after an integration by parts using the identity 
(\ref{idd8}), and taking into account Eq. (\ref{lEqmi}), 
the variation (\ref{vSsf3}) of the superparticle action reads 
\begin{eqnarray}
\label{vSsf5}
\delta^\prime S_{sp} & 
= \int d^8 Z E  \{ i {\cal D}_{\alpha} {\cal K}_a^{\alpha} - 
i \bar{{\cal D}}_{\dot{\alpha}} \bar{{\cal K}}_a^{\dot{\alpha}} 
\qquad \nonumber \\ & 
\qquad - {1\over 4} \tilde{\sigma}_b^{ \dot{\alpha} {\alpha} }
[{\cal D}_{{\alpha}}, \bar{{\cal D}}_{\dot{\alpha}}] {\cal K}_a{}^b\}
\delta H^a  \; , 
\end{eqnarray} 
where the `spin 3/2' and `spin 2'  `current prepotentials',  
${\cal K}_a^{\alpha}$, $\bar{{\cal K}}_a^{\dot{\alpha}}= 
({\cal K}_a^{\alpha})^*$ and  ${\cal K}_a{}^b$, are defined by 
\begin{eqnarray}
\label{K3/2}
& {\cal K}_a^{\alpha}(Z):= \int_{W^1}  
 {l(\tau )\over \hat{E}} \hat{E}_{a\tau} \hat{E}^{\alpha} 
\delta^8(Z-\hat{Z})  \; , \\ \label{bK3/2} 
& \bar{{\cal K}}_a^{\dot{\alpha}}(Z):=
\int_{W^1}  
 {l(\tau )\over \hat{E}} \hat{E}_{a\tau} \hat{E}^{\dot{\alpha}}
\delta^8(Z-\hat{Z}) 
\; , \\ \label{K2}
& {\cal K}_a{}^b(Z):= \int_{W^1}  
 {l(\tau )\over \hat{E}} \hat{E}_{a\tau} \hat{E}^{b}
\delta^8(Z-\hat{Z}) \;  .
\end{eqnarray}

Eqs. (\ref{vSsf5}) and (\ref{vSsf}) imply the appearance of a  
current potential superfield ${\cal J}_a$ (which is 
a vector density distribution 
with support on the worldline), 
in the vector superfield equation of the coupled system 
({\it cf.}  (\ref{SGeqmG})), 
\begin{eqnarray}
\label{Ga=Ja}
 {\delta S\over \delta H^a}=0 \quad \Rightarrow \quad  G_a ={\cal J}_a \; .  
\end{eqnarray} 
This vector current potential is constructed from the vector--spinor 
and tensor densities 
Eq. (\ref{K3/2}), (\ref{bK3/2}), (\ref{K2}) 
(hence the `current prepotential' name for ${\cal K}_a^{\;B}$) 
as follows 
\begin{eqnarray}\label{Ja=}
& {1\over 6}{\cal J}_a = - i {\cal D}_{\alpha} {\cal K}_a^{\alpha} 
+ i \bar{{\cal D}}_{\dot{\alpha}} \bar{{\cal K}}_a^{\dot{\alpha}} 
+ {1\over 4} \tilde{\sigma}_b^{ \dot{\alpha} {\alpha} }
[{\cal D}_{{\alpha}}, \bar{{\cal D}}_{\dot{\alpha}}] {\cal K}_a{}^b \; .
\end{eqnarray}

The preservation of the scalar superfield equation 
$R=0$ in the interacting dynamical system (\ref{SGSP}), 
Eq. (\ref{SeqmbR}),   immediately 
implies the vanishing of the spin $1/2$ part of the 
superfield generalization of the gravitino field strength, 
\begin{eqnarray}\label{SGRS+}
& (\sigma^a \tilde{\sigma}^b)_\beta{}^\gamma T_{ab\gamma}= 0\; ,
\qquad (\tilde{\sigma}^a\sigma^b)^{\dot{\gamma}}{}_{\dot{\beta}} 
T_{ab\dot{\gamma}}= 0\; 
\end{eqnarray}
(see Eq. (\ref{gr1}). 
However, the above equation is only a part of the content 
of the superfield generalization of the free Rarita-Schwinger 
equation\footnote{Indeed, the linear approximation equation  
$\epsilon^{abcd}\partial_b\psi_c^{\alpha}\sigma_{d\alpha\dot{\alpha}}=0$   
is equivalent to the equations 
$(\sigma^a \tilde{\sigma}^b)_{(\beta\gamma)} \partial_b\psi_a^{\gamma}=0$ 
(which is a counterpart of (\ref{SGRS+})) 
{\sl and} $ \partial^c\psi_c^{\alpha}=0$.}. 
The complete {\sl superfield generalization of the 
Rarita-Schwinger equation} for the coupling system 
can be obtained from Eq. (\ref{Tabg}) with $R=0, G_a={\cal J}_a$ and 
possesses the source term 
\begin{eqnarray}\label{SGRS=J}
\epsilon^{abcd}T_{bc}{}^{\alpha}\sigma_{d\alpha\dot{\alpha}}= 
{i\over 8} \tilde{\sigma}^{a\dot{\beta}\beta} \bar{{\cal D}}_{(\dot{\beta}|}
{\cal J}_{\beta|\dot{\alpha})}   \; .
\end{eqnarray}

Using (\ref{Rscalar}) and Eq. (\ref{SeqmbR}), $R=0$, 
one finds that the superfield generalization 
of the scalar curvature vanishes in the supergravity--superparticle 
interacting system, 
\begin{eqnarray}
\label{Rscalar=0}  
R_{ab}{}^{ab}= 0\; .
\end{eqnarray}
However, in accordance with Eq. (\ref{RRici}), 
the Ricci tensor is expressed not only through 
$R$, $\bar{R}$, but also through the $G_a$ superfield. 
Hence in the interacting system the {\sl superfield 
Einstein equation} acquires a source term which is 
expressed through a second derivative of the current 
potential superfield
${\cal J}^{{\alpha}\dot{\beta}}=
{\cal J}^a\tilde{\sigma}_a^{{\alpha}\dot{\beta}}$:  
\begin{eqnarray}
\label{EiInt}  
R_{bc}{}^{ac} & = 
{1\over 32} ({{\cal D}}^{{\beta}}\bar{{\cal D}}^{(\dot{\alpha}|} 
{\cal J}^{{\alpha}|\dot{\beta})} - 
\bar{{\cal D}}^{\dot{\beta}} {{\cal D}}^{({\beta}}
{\cal J}^{{\alpha})\dot{\alpha}})
\sigma^a_{\alpha\dot{\alpha}}\sigma_{b\beta\dot{\beta}} \;  .
\hspace{-1cm}
\nonumber \\  
\end{eqnarray}

\subsection{Properties of current potential ${\cal J}_a$  
and ${\cal K}_a{}^B$ prepotentials}

Thus the vector superfield supergravity 
equation acquires the 
source 
(\ref{Ja=}) (`current potential') 
from the Brink--Schwarz superparticle action $S_{sp}$, Eq. 
(\ref{Ga=Ja}), while the scalar superfield equations 
(\ref{SeqmbR}), (\ref{SeqmR}) remain sourceless as 
in free supergravity. 
Then the identities (\ref{DG=DR}) immediately result in 
\begin{eqnarray}\label{DJa=0}
& {\cal D}^{\alpha} {\cal J}_{\alpha\dot{\alpha}} = 0 \; , \qquad 
\bar{{\cal D}}^{\dot{\alpha}}{\cal J}_{\alpha\dot{\alpha}} = 0 \; , 
\end{eqnarray}
which imply  the supercurrent conservation 
\begin{eqnarray}\label{DaJa=0}
& {\cal D}^{a} {\cal J}_{a} = 0 \; . \qquad 
\end{eqnarray}

In accordance with Eq. (\ref{Ja=}), 
the superparticle current is constructed from the current prepotentials 
(\ref{K3/2}), (\ref{bK3/2}), (\ref{K2}). Moreover, 
 Eq. (\ref{Ja=}) can be presented in the form  
\begin{eqnarray}\label{Ja=2}
{1\over 6} {\cal J}_a & = - i {\cal D}_{\alpha} 
({\cal K}_a^{\alpha} 
+ {i\over 4} \tilde{\sigma}_b^{ \dot{\alpha} {\alpha} }
\bar{{\cal D}}_{\dot{\alpha}} {\cal K}_a{}^b)  + 
\nonumber \\ & + i \bar{{\cal D}}_{\dot{\alpha}} 
(\bar{{\cal K}}_a^{\dot{\alpha}} + {i\over 4} 
\tilde{\sigma}_b^{ \dot{\alpha} {\alpha} }
{\cal D}_{{\alpha}} {\cal K}_a{}^b )\; .   
\end{eqnarray}

It is interesting that the spinor--vector and tensor 
current prepotential carry only the irreducible spin $3/2$ and spin $2$ 
representation of the Lorentz group, respectively.
Indeed, extracting the worldline measure $d\tau$ in (\ref{K2}), 
 ${\cal K}^{ab}(Z)= \int_{W^1}  d\tau 
 {l(\tau )\over \hat{E}} \hat{E}^a_{\tau} \hat{E}_{\tau}^{b}
\delta^8(Z-\hat{Z})$, one easily sees that the tensor ${\cal K}^{ab}(Z)$ 
is symmetric, and traceless 
due to Eq. (\ref{lEqmi}). 
In this sense one can say that the current potential 
contains only a spin $2$ irreducible part, 
\begin{eqnarray}
\label{K2+}
 {\cal K}^{ab}(Z)= {\cal K}^{ba}(Z) \; , \qquad  {\cal K}_b{}^b(Z)=0 \; .
\end{eqnarray}
The spinor--vector current prepotentials 
carry spin $3/2$, because, due to Eq. (\ref{fEqm}), their spin $1/2$ 
irreducible parts vanish, 
\begin{eqnarray}
\label{K3/2+}
 {\cal K}_a^{\alpha}(Z) \sigma^a_{\alpha \dot{\alpha}}
\equiv {\cal K}_{\alpha \dot{\alpha}}{}^{\alpha} =0\; , 
\qquad 
\\ \label{bK3/2+}
\sigma^a_{\alpha \dot{\alpha}} 
\bar{{\cal K}}_a^{\dot\alpha}(Z) \equiv  
\bar{{\cal K}}_{\alpha \dot{\alpha}}{}^{\dot\alpha}(Z) =0\; . 
\qquad 
\end{eqnarray}
Finally, using Eq. (\ref{bEqm}), 
together with the identities 
\\ $ E_A{}^M(Z)\partial_M \delta(Z-\hat{Z}) =
E_A{}^M(\hat{Z})\partial_M \delta(Z-\hat{Z}) -
 (-1)^{M+AM}\partial_ME_A{}^M({Z})\delta(Z-\hat{Z})\,$ , 
$\; \partial_M \delta(Z-\hat{Z})= -
\partial/\partial \hat{Z}^M\,  \delta(Z-\hat{Z})\;$ and 
$\; (-1)^{M+AM} {\cal D}_M(EE_A{}^M)\equiv E (-1)^B T_{AB}{}^B=0$
(the last part of the last identity is valid due to the 
supergravity constraints), 
one finds the relation
\begin{eqnarray}
\label{DK}
(-)^B {\cal D}_B{\cal K}_a^B \equiv 
{\cal D}_b{\cal K}_a{}^b - {\cal D}_{\beta}{\cal K}_a^{\beta} - 
\bar{{\cal D}}_{\dot\beta}\bar{{\cal K}}_a^{\dot\beta} =0\; , 
\end{eqnarray}
which completes the list of the properties of the 
superparticle current prepotentials (\ref{K3/2}), (\ref{bK3/2}), (\ref{K2}).

Contracting the vector indices of the current prepotentials 
with the $\sigma$ matrices, one can write the irreducibility 
conditions (\ref{K2+}), (\ref{K3/2+}), (\ref{bK3/2+}) in the form 
\begin{eqnarray}
\label{K2+s}
{\cal K}^{\alpha \beta \; \dot{\alpha}\dot{\beta}}\equiv  
 {\cal K}^{ab}\tilde{\sigma}_a^{\dot{\alpha}\alpha}
\tilde{\sigma}_b^{\dot{\beta}\beta } = 
{\cal K}^{(\alpha \beta) \; (\dot{\alpha}\dot{\beta})} \; ,
\\ \label{K3/2+s}
{\cal K}^{\alpha \beta \dot{\beta}}
\equiv 
 {\cal K}_a^{\alpha} \tilde{\sigma}_a^{\dot{\beta}\beta}
={\cal K}^{(\alpha \beta)\dot{\alpha}}\; , 
\\ \label{bK3/2+s}
\bar{{\cal K}}^{\dot{\alpha}\dot{\beta}\beta }
\equiv 
 \bar{{\cal K}}_a^{\dot\alpha} \tilde{\sigma}_a^{\dot{\beta}\beta}
= \bar{{\cal K}}^{(\dot{\alpha}\dot{\beta})\beta } \; .
\end{eqnarray}
Then, relation (\ref{DK}) reads 
\begin{eqnarray}
\label{DK1}
{1\over 2} {\cal D}_{\beta\dot{\beta}}
{\cal K}^{\alpha \beta \; \dot{\alpha}\dot{\beta}} - 
{\cal D}_{\beta}{\cal K}^{\alpha \beta \dot{\alpha}}
- \bar{{\cal D}}_{\dot{\beta}}\bar{{\cal K}}^{\dot{\alpha}\dot{\beta}\alpha }
=0\; , 
\end{eqnarray}
or, equivalently, 
\begin{eqnarray}
\label{DK2}
& {\cal D}_{\beta}
({\cal K}^{\alpha \beta \dot{\alpha}} + 
{i\over 4} \bar{{\cal D}}_{\dot{\beta}} 
{\cal K}^{\alpha \beta \; \dot{\alpha}\dot{\beta}})= 
- \bar{{\cal D}}_{\dot{\beta}}
(\bar{{\cal K}}^{\dot{\alpha}\dot{\beta}\alpha }
+ {i\over 4} {\cal D}_{\beta}
{\cal K}^{\alpha \beta \; \dot{\alpha}\dot{\beta}})
\; . \hspace{-1.0cm} \nonumber \\ 
\end{eqnarray}

Eqs. (\ref{DK1}), (\ref{DK2})  allow us to write the 
expression (\ref{Ja=2}) for supercurrent 
in two other equivalent forms 
\begin{eqnarray}
{1\over 6} {\cal J}^{\alpha\dot{\alpha}} \equiv & {1\over 6}{\cal J}_a 
\tilde{\sigma}^{a \dot{\alpha} {\alpha} }= 
\hspace{3.5cm}
\nonumber \\  
\label{Ja=4}
& = - 2i {\cal D}_{\beta} 
({\cal K}^{(\beta\alpha )\; \dot{\alpha}} 
+ {i\over 4}
\bar{{\cal D}}_{\dot{\beta}} 
{\cal K}^{(\alpha \beta) \; (\dot{\alpha}\dot{\beta})})
\\ \label{Ja=5}
& =  2i \bar{{\cal D}}_{\dot{\alpha}} 
(\bar{{\cal K}}^{(\dot{\alpha}\dot{\beta})\beta }
+ {i\over 4} 
{\cal D}_{{\beta}} {\cal K}^{(\alpha \beta) \; (\dot{\alpha}\dot{\beta})}
 )\; . 
\end{eqnarray}
Now one can easily derive 
Eq. (\ref{DJa=0}) 
using (\ref{Ja=4}) and  (\ref{Ja=5}). To this aim one uses  
the algebra of spinor derivatives of the same chirality, 
Eq. (\ref{(DD)1}),  
\begin{eqnarray}
\label{(DD)2}
R=\bar{R}=0 \quad  \Rightarrow \quad 
\{ {\cal D}_\alpha , {\cal D}_\beta \} = 0 \; . 
\end{eqnarray}
Then the current potential conservation, Eq. (\ref{DaJa=0}), 
follows from Eqs. (\ref{DJa=0}) and 
Eq. (\ref{(DbD)}).

The properties (\ref{DJa=0}) imply  
$\bar{{\cal D}}_{\dot{\alpha}}
{\cal J}_{\beta\dot{\beta}}
=
\bar{{\cal D}}_{(\dot{\beta}|} {\cal J}_{\beta|\dot{\alpha})}$ and, 
hence, 
 allow us to write the {\sl r.h.s} 
of the {\sl superfield Rarita--Schwinger equation} 
as the fermionic covariant  derivative of the current potential 
\begin{eqnarray}\label{SGRS=J1}
\Psi^a_{\dot\alpha} \equiv 
\epsilon^{abcd}T_{bc}{}^{\alpha}\sigma_{d\alpha\dot{\alpha}}= 
{i\over 4} \bar{{\cal D}}_{\dot{\alpha}}
{\cal J}^a   \; .
\end{eqnarray} 
The {\sl superfield Einstein equation} (\ref{EiInt}) can be written as
\begin{eqnarray}\label{EiInt1} 
R_{bc}{}^{ac}& = {1\over 16} 
\tilde{\sigma}_b^{\dot{\beta}\beta} [{\cal D}_{\beta}, 
\bar{{\cal D}}_{\dot{\beta}}] {\cal J}^a
\end{eqnarray}

Eqs. (\ref{SGRS=J1}), (\ref{EiInt1}) exhibit an 
 interdependence of the 
Einstein and Rarita--Schwinger superfield equations, 
\begin{eqnarray}\label{EiRS} 
R_{bc}{}^{ac}& = -{i\over 4} \tilde{\sigma}_b^{\dot{\beta}\beta}
({\cal D}_\beta \Psi^a_{\dot{\beta}} + 
\bar{{\cal D}}_{\dot{\beta}} \bar{\Psi}^a_\beta) \; .
\end{eqnarray}

\section{Gauge fixing and equations of motion 
for the interacting system}

\subsection{Gauge fixing}

As all the gauge symmetries of the `free' superfield supergravity 
are still present in the interacting system, one can fix first the WZ 
gauge (\ref{WZgauge}). 
This would be the first step towards the component description of the 
interacting system
in terms of the  usual graviton and gravitino spacetime fields. 

As  was shown in Sec. IIB and in Appendix C2,  the WZ 
gauge is preserved by some specific superdiffeomorphisms and 
superspace local Lorentz transformations with  
free  parameters 
$b_0^A(x)= (b_0^a (x), \varepsilon_0^{\underline{\alpha}}(x))= 
b^A(Z)\vert_{\theta=0}$ and $l^{ab}(x)=L^{ab}(Z)\vert_{\theta=0}$. 
In accordance with 
Eqs. (\ref{sdiffhZ}) and (\ref{4kappa}),  the transformation  
of the fermionic coordinate function 
$\hat{\theta}^{\breve\alpha}(\tau)$ 
under superdiffeomorphisms 
and  worldline $\kappa$--transformations 
acquires the form 
\begin{eqnarray}\label{thG0}
\delta  \hat{\theta}^{\breve\alpha}(\tau)=  
b^{\breve\alpha}(\hat{Z}) 
+ \delta_{\kappa}  \hat{\theta}^{\breve\alpha}(\tau)
\; , 
\end{eqnarray} 
where $\delta_{\kappa}  \hat{\theta}^{\breve\alpha}(\tau)$ is defined by the 
Eq. (\ref{4kappa}) with 
$M= \breve{\alpha}$. 
This transformation rule reflects the Goldstone nature of the 
superparticle (or superbrane) coordinate function 
\cite{Polchinski} (see also \cite{Ivanov,Pashnev} and refs. therein). 

In the WZ gauge (\ref{WZgauge}), Eq. (\ref{thG0}) can be written in the form 
(see Eqs. (\ref{WZgth}), (\ref{bdecWZ}))
\begin{eqnarray}\label{thG1}
& \delta  \hat{\theta}^{\underline{\alpha}}(\tau)
\equiv  \delta  \hat{\theta}^{\breve\alpha}(\tau) 
\delta_{\breve\alpha}{}^{\underline{\alpha}} = 
\varepsilon^{\underline{\alpha}}(\hat{Z}) +
\delta_{\kappa} 
\hat{\theta}^{\underline{\alpha}}(\tau)\; .\hspace{-0.2cm}
\end{eqnarray} 

Decomposing the {\it r.h.s} of Eq. (\ref{thG1}) in power series in 
$\hat{\theta}(\tau)$ one writes 
\begin{eqnarray}\label{thG}
\delta  \hat{\theta}^{\underline{\alpha}}(\tau) & = 
\varepsilon^{\underline{\alpha}}(\hat{Z}) 
\vert_{\hat{\theta}=0} + 
\delta_{\kappa}  \hat{\theta}^{\underline\alpha}(\tau)\vert_{\hat{\theta}=0} + 
{\cal O} (\hat{\theta}) = \nonumber \hspace{-1.5cm}
\\ &
= \varepsilon_0^{\underline{\alpha}}(\hat{x}) + 
\delta_{\kappa}  \hat{\theta}^{\underline\alpha}(\tau)\vert_{\hat{\theta}=0}
+ 
{\cal O} (\hat{\theta})\; , 
\end{eqnarray} 
where the arbitrary fermionic field parameter 
$\varepsilon_0^{\underline{\alpha}}(\hat{x})$ is defined as in 
Eqs. (\ref{bdecWZ}), (\ref{WZgrs2s}) 
 and {\sl corresponds to one of the  symmetries 
that preserve the WZ gauge}. 

{\sl Thus we can fix} {\sl the gauge} 
(simultaneously with the WZ gauge) 
\begin{eqnarray}\label{thGAUGE}
\hat{\theta}^{\underline{\alpha}}(\tau) = 0 \;  
\end{eqnarray}
({\it cf.} the description of super--Higgs effect in \cite{Volkov}) 
{\sl by using the freedom in the fermionic parameters 
$\varepsilon_0^{\underline{\alpha}}(\hat{x})$} (but 
{\sl not} the pull--back 
$\varepsilon_0^{\underline{\alpha}}(\hat{x},\hat{\theta})$
of the complete superfield 
$\varepsilon_0^{\underline{\alpha}}(Z)$). 
The 
gauge (\ref{thGAUGE}) is preserved by transformations such that 
\begin{eqnarray}\label{thGpres}
\varepsilon_0^{\underline{\alpha}}(\hat{x})= 
- \delta_{\kappa} 
\hat{\theta}^{\underline{\alpha}}(\tau)\vert_{\hat\theta=0}
= - \kappa^{\underline{\beta}}
\gamma_a{}_{\underline{\beta}}{}^{\underline{\alpha}}
\hat{E}_\tau^a \vert_{\hat\theta=0}\; ,
\end{eqnarray} 
where we have written the form of the $\kappa$--symmetry transformations 
(\ref{4kappa}) explicitly, in Majorana spinor notation,    
by using the WZ gauge relations.

\subsection{On the Goldstone nature of the (super)brane coordinate 
functions}

Since the possibility of fixing the gauge (\ref{thGAUGE})
might look unexpected, we now discuss  its physical 
meaning. 

First, let us note that similar considerations show that the 
bosonic counterpart of the gauge (\ref{thGAUGE}) can also be 
fixed on the bosonic coordinate functions. It reads
\footnote{Note that the gauge with all components of 
$\hat{x}^{\mu}(\tau)$ equal to zero cannot be fixed due to the 
restrictions on the pure bosonic sector of the 
transformations since the diffeomorphism transformations have 
to be invertible and  it is clear that a (world)line could not be 
represented by one point in any nondegenerate coordinate system.
In contrast, the nondegeneracy of superdiffeomorphisms 
implies $det (\delta_{\check\alpha}{}^{\check\beta} + 
{\partial b^{\check\beta}(x,\theta) \over 
\partial \theta^{\check\alpha}})\not=0$, which does not 
restrict the field parameter $b^{\check\beta}(x,0)$ and, hence, 
$\varepsilon_0^{\underline{\alpha}}(x)$ (see (\ref{bdecWZ})) in 
 (\ref{thG}). This allows us to use the pull--back  
$\hat{\varepsilon}_0^{\underline{\alpha}}
:= \varepsilon_0^{\underline{\alpha}}(\hat{x})$
 of $\varepsilon_0^{\underline{\alpha}}(x)$
to fix 
the gauge (\ref{thGAUGE}), where all the components of 
$\hat{\theta}^{\check{\alpha}}(\tau)$ are set to zero.}
\begin{eqnarray}\label{xGAUGE}
\hat{x}^{\mu}(\tau) = (\tau, 0, 0, 0) \;   
\end{eqnarray}
(or $x^{\mu}(\tau) = (\tau, 0, 0, \pm \tau)$ if one identifies 
$x^0$ with the time--like dimension in the flat (super)space limit). 
In general, for a $D$ dimensional $p$--brane interacting with dynamical 
gravity one can fix {\it locally}  
 the following counterpart of the gauge (\ref{xGAUGE}) (static gauge)  
\begin{eqnarray}\label{xGAUGEp}
\hat{x}^{\mu}(\tau, \vec{\sigma}) = 
(\tau, \sigma^1, \ldots , \sigma^p, 0, \ldots , 0) \; , 
 \end{eqnarray}
where the first $(p+1)$ of the $D$ coordinate functions are identified with 
the local worldvolume coordinates 
$\xi^m = (\tau, \sigma^1, \ldots , \sigma^p)$, and the remaining 
coordinate functions are set to zero (see \cite{BdAI1}). 

Clearly, the gauge (\ref{xGAUGE}) or (\ref{xGAUGEp}) 
can be fixed also in a dynamical system of pure bosonic gravity 
interacting with a bosonic particle or brane.  
As such, this phenomena should be known in general relativity, and 
this is indeed the case. The pure gauge nature of the 
coordinate functions describing the motion of a dynamical  
source (particle) was already known in general relativity,  see {\it e.g.} 
\cite{Inf,Fock}. 
The presence of the gauge symmetry allowing  to fix 
{\sl locally} the gauge (\ref{xGAUGEp}) for branes or 
(\ref{xGAUGE}) for a particle is reflected in the language of second Noether 
theorem (see \cite{BdAI1,BdAI2}) by stating that the brane or 
particle equations of motion can be derived as a consequence of 
the field equations for gravity. 
This type of statement can be found in books 
(see, {\it e.g.} p. 240 in \cite{Fock},  pp. 19, 44-48 and 
Eq. (1.6.13) in  \cite{Inf}) and 
comes back to the original paper by Einstein and Grommer \cite{EiGro}. 
Namely, one can derive the equations of motion of the particle source 
from the covariant conservation of the particle  energy--momentum tensor 
 in the {\it r.h.s.} of the Einstein field equation. 
So, the statement of \cite{Inf} is that we do not need varying the 
action with respect to the matter (particle) variables, 
because we can obtain the equations of motion for the matter part  
as a consequence of the Einstein equations. These, by their geometric   
structure, imply the covariant energy--momentum conservation which in turn 
is equivalent to the matter equations of motion.

Clearly, for the case of the brane source the same arguments result in 
the derivation of the equations of motion for the brane variables 
from the conservation of an energy--momentum tensor with 
support on worldvolume  (see \cite{BdAI1} for an explicit  proof). 
Then the choice of {\sl local} coordinate system allows one to 
fix the gauge (\ref{xGAUGEp}) {\sl locally}. Certainly, for 
topologically nontrivial and/or closed worldvolume this 
gauge cannot be fixed globally. In contrast, one immediately 
notices that there are no restrictions on a  global fixing of 
the fermionic gauge  (\ref{thGAUGE}) as no way of introducing 
topology on a Grassmann algebra is known.

Thus one can state that both the fermionic and bosonic coordinate functions 
of superbrane are pure gauge (can be gauged away) when the interacting system 
of {\sl dynamical} ({\sl not}  background) {\sl supergravity} 
and a dynamical superbrane is considered. 

Actually, the above statement is tantamount to saying that
the coordinate functions of superbranes are 
{\sl Goldstone fields}\footnote{
More precisely, the bosonic and fermionic 
Goldstone fields are identified, respectively,  
with the bosonic coordinate functions corresponding to the directions 
orthogonal to the worldvolume and with a half of  fermionic  coordinate 
functions.}. 
In flat superspace these Goldstone fields correspond to 
the translational symmetry and global supersymmetry that are broken by 
the superbrane worldvolume \cite{Polchinski,Ivanov}  
{\it i.e.}, by the position of the superbrane in superspace.
Then, when a brane or particle interacting with (super)gravity 
is considered and, {\sl moreover}, (super)gravity is described 
by an action on the same footing as the (super)brane, 
the global translations and global supersymmetry are replaced by 
{\sl superdiffeomorphism} symmetry, which is the {\sl gauge} 
symmetry of the coupled action ({\it e.g.} of the action (\ref{SGSP}); 
see also Appendix B2).  
Thus the coordinate functions in such a dynamical 
system should be considered as {\sl Goldstone fields for gauge 
symmetries}.

The Goldstone fields for the  {\sl gauge} symmetries are {\sl always} 
pure gauge fields (compensators in the supergravity language).   
The `unitary' gauge where the Goldstone degrees of freedom are set to zero 
is always assumed in the consideration of Higgs phenomenon. 
For the case of spontaneously broken {\sl internal} gauge symmetry, 
the only trace of the interaction with the Goldstone fields 
in this gauge turns out to be
{\sl the  mass terms} in the gauge field equations. 
This is just the content of the standard Higgs phenomenon.  

Now, when the Goldstone fields for {\sl spacetime} 
(or superspace) gauge symmetry 
live on a subspace of spacetime (superspace), {\it i.e.} on the (super)brane 
worldvolume or (super)particle worldline, we may also expect a modification 
of the equations for the spacetime (or superspace) gauge fields. 
However,  such a modification will only be produced by terms with support 
on the 
worldvolume or worldline. Hence these  new terms modifying the gauge field 
equations should be just  the {\sl 
source terms}, like the {\it r.h.s.} of Eq. (\ref{EiInt02}) 
below (in particular, for $\hat{x}$ given by 
(\ref{xGAUGE})). Summarizing, {\sl when the Goldstone fields are 
worldvolume fields, 
the counterpart of the mass terms appearing in the gauge field 
equations as a result of the usual Higgs mechanism, are precisely the source 
terms in the Einstein equation and in 
some other gauge (super)field equations}. 

In complete correspondence with the usual Higgs phenomenon, the bosonic 
`unitary' gauge (\ref{xGAUGEp}) clearly cannot remove the source from the 
Einstein equation. 
However, 
the  super--Higgs effect \cite{Volkov} 
may be subtler when we have {\sl fermionic} 
Goldstone fields defined on a 
surface in superspace ({\it i.e.} on the superbrane worldvolume). 
The gauge field equations that acquire a source term as a result of 
the super--Higgs effect would be the {\sl superfield generalizations} of the 
Einstein equations and other gauge field equations, including that for 
the gravitino, $\Psi=J_{\Psi}$ (see  (\ref{SGRS=J1})). 
Let us discuss the fermionic {\sl superfield} source term $J_{\Psi}$.  
In the `unitary' gauge $\hat{\theta}^{\check\alpha}(\xi)=0$ 
one can expect that $J_{\Psi}\propto \theta$ (we  show below 
that this is indeed the case for $D=4$, $N=1$  
supergravity--superparticle interacting system). 
Now let us recall that the spacetime fermionic gauge field equation 
(the gravitino equation) 
is given by the {\sl leading component} of the superfield 
equation $\Psi=J_{\Psi}$, {\it i.e.} by 
$\Psi\vert_{\theta =0} = J_{\Psi}\vert_{\theta =0}$. Thus, if 
$J_{\Psi}\propto \theta$, this gives 
  $J_{\Psi}\vert_{\theta =0}=0$. This means that the {\sl spacetime} 
equation for fermionic gauge field becomes  sourceless, 
$\Psi\vert_{\theta =0} =0$, in the `unitary' gauge  
$\hat{\theta}^{\check\alpha}(\xi)=0$ 
(Eq. (\ref{thGAUGE}) for the superparticle).

We hope to return to the discussion of the fate of the superbrane 
degrees of freedom and other issues of the (super)Higgs 
phenomenon in the presence of superbranes in a future publication. 
Here our goal is more immediate: to find the explicit form of the 
equations of motion of the supergravity--superparticle interacting system 
in the fermionic `unitary' gauge (\ref{thGAUGE}).

\subsection{Gauge fixed form of the equations of motion 
of the coupled system}

In the WZ gauge supplemented by the condition 
(\ref{thGAUGE}), {\sl the  coupled system action is reduced to the action for 
supergravity interacting with a bosonic particle}. 
After integration on Grassmann variable $\theta$ in the supergravity 
part of the coupled action (\ref{SGSP}) this coupled action
should become basically the same as the $D=4$ case of the action for 
{\sl supergravity--bosonic particle} interacting  
system considered in ref. \cite{BdAI1}. 
The only expected difference is   the presence of the auxiliary fields, 
$G_a\vert_{\theta=0}$, $R\vert_{\theta=0}$, $\bar{R}\vert_{\theta=0}$,  
which are not essential 
as they appear in the component  action 
only through  quadratic combinations, 
without derivatives \cite{Siegel79} and can be removed using  their 
algebraic equations of motion. 
Furthermore, passing 
to the component approach to supergravity, which  
 deals with fields on  spacetime, one excludes 
the superspace diffeomorphisms 
$\delta_{diff} (b^M)$ with 
$\theta^{\breve\alpha} \rightarrow 
\theta^{\breve\alpha} + b^{\breve\alpha}(x, \theta)$ from consideration.
Then Eq. (\ref{thGpres}) is treated as the partial 
breaking of the local 
{\sl spacetime} supersymmetry  \cite{BdAI1}
(originating in $\tilde{\delta}_{gc}$ and given by 
Eqs. (\ref{lsusyEa}), (\ref{lsusyEal}), (\ref{lsusyEbal})
with $\theta=0$ and $G_a\vert_{\theta=0}=0= R\vert_{\theta=0}$).

Having in mind the results of \cite{BdAI1}, 
one would expect that, in the light of above correspondence,  
the auxiliary fields should  have  
vanishing values  in the gauge (\ref{thGAUGE}) and  
that the spacetime Rarita--Schwinger equations 
following from the superfield action for the interacting system, 
Eq. (\ref{SGSP}), would be sourceless in this gauge.

The analysis indicates that this is indeed the case. 
Firstly, in the coupled system the scalar main superfields 
(\ref{4WR}) 
are equal to zero 
on the mass shell, $R=0=\bar{R}$, Eqs. (\ref{SeqmbR}), (\ref{SeqmR}). 
Thus $R\vert_{\theta=0}=0, \bar{R}\vert_{\theta=0}=0$. 
In contrast,  the vector main superfield (\ref{4WGa}) becomes equal 
to the current potential 
(\ref{Ja=}), Eq. (\ref{Ga=Ja}). Hence $G_a\vert_{\theta=0}=
{\cal J}_a\vert_{\theta=0}$. However, it is seen that 
${\cal J}_a\vert_{\theta=0}=0$  in the gauge (\ref{thGAUGE}). 
Indeed, ${\cal J}_a$ is constructed from the current prepotentials 
(\ref{K3/2}), (\ref{bK3/2}), (\ref{K2}), which involve 
$\delta^8(Z-\hat{Z}) \equiv 
(\theta-\hat{\theta}(\tau))^4 
\delta^4(x-\hat{x})$, Eqs. (\ref{delta}), (\ref{delta8}). 
In the gauge 
(\ref{thGAUGE})
\begin{eqnarray}
\label{K3/20}
& {\cal K}_a^{\alpha}(Z)= 
(\theta)^4 
\int_{W^1}  
 l(\tau ) [{1\over \hat{E}} \hat{E}_{a\tau} \hat{E}^{\alpha}
] \vert_{\hat{\theta}=0} 
\delta^4(x-\hat{x})
\; , 
\\ 
\label{bK3/20} 
& \bar{{\cal K}}_a^{\dot{\alpha}}(Z)=
(\theta)^4 \int_{W^1}  
 l(\tau )[{1\over \hat{E}} \hat{E}_{a\tau} \hat{E}^{\dot{\alpha}}
] \vert_{\hat{\theta}=0} 
\delta^4(x-\hat{x}) 
\; , \\ \label{K20}
& {\cal K}_a{}^b(Z)= (\theta)^4  \int_{W^1}  
 l(\tau )[{1\over \hat{E}} \hat{E}_{a\tau} \hat{E}^{b}
] \vert_{\hat{\theta}=0} 
\delta^4(x-\hat{x}) \;  ,
\end{eqnarray}
{\it i.e.} all current prepotentials become proportional to the highest 
possible power in the superspace Grassmann coordinates, 
\begin{eqnarray} \label{K3/201}
\hat{\theta}=0 \;  \qquad \Rightarrow  
\qquad  
\cases{ {\cal K}_a{}^{\beta}(Z)\propto
(\theta)^4\; , 
\cr  
 \bar{{\cal K}}_a{}^{\dot{\beta}}(Z)
\propto
(\theta)^4 \; , \cr 
 {\cal K}_a{}^b(Z)\; 
\propto
(\theta)^4 .}
\end{eqnarray}
Thus only the action of {\sl four} Grassmann covariant derivatives 
on  ${\cal K}_a^{A}(Z)=({\cal K}_a{}^b\, ,
{\cal K}_a^{\alpha}\, , 
 \bar{{\cal K}}_a^{\dot{\alpha}})$ can produce an expression which has 
a nonvanishing value for $\theta=0$. 
In particular, 
\begin{eqnarray} \label{Ja=th3}
\hat{\theta}=0 \;  
\qquad \Rightarrow \qquad 
{\cal J}_a \propto (\theta)^2 \; , 
\end{eqnarray}
and, hence, the auxiliary vector field of the minimal $D=4$, $N=1$ 
supergravity vanishes on the mass shell in the gauge (\ref{thGAUGE}), 
\begin{eqnarray} \label{Ja0=0}
\hat{\theta}=0 \; , 
\quad \Rightarrow \quad 
G_a\vert_{\theta =0}= 
{\cal J}_a \vert_{\theta =0}=0\; . 
\end{eqnarray}
 
The Rarita--Schwinger equation can be derived setting 
$\theta=0$ in the superfield equation (\ref{SGRS=J1}). 
However, in accordance with Eq. (\ref{K3/201}), 
${\cal D}_A {\cal J}_a \vert_{\theta=0}=0$. Hence {\sl the spacetime
Rarita--Schwinger equation derived from the superfield action 
for interacting supergravity--superbrane system becomes sourceless
in the gauge (\ref{thGAUGE})}, 
\begin{eqnarray}\label{SGRS=0}
\hat{\theta}=0 \;  
\quad \Rightarrow \hspace{3cm} \nonumber \\ 
\Psi^a_{\dot\alpha} \vert_{\theta=0}\equiv 
\epsilon^{abcd}T_{bc}{}^{\alpha}\sigma_{d\alpha\dot{\alpha}}\vert_{\theta=0}= 
{i\over 4} \bar{{\cal D}}_{\dot{\alpha}}
{\cal J}^a  \vert_{\theta=0}=0 \; .
\end{eqnarray} 
One can verify using Eq. (\ref{TbbfWZ}) that, due to 
Eqs. (\ref{Ja0=0}), (\ref{SeqmbR}), 
$T_{ab}{}^{\alpha}\vert_{\theta =0} = 2 e_a^\mu e_b^\nu 
{\cal D}_{[\mu}\psi_{\nu]}^{\alpha}(x)$. Hence 
the above  statement is related to the true component gravitino equation.

The component Einstein equation for the coupled system can be obtained
by setting  $\theta=0$ in Eq. (\ref{EiInt1}). Clearly, it possesses 
 a  source term, but only from the spin 2 current prepotential, 
Eq. (\ref{K2}),  
\begin{eqnarray}\label{EiInt01} 
& R_{bc}{}^{ac} \vert_{\theta=0}  = {1\over 16} 
\tilde{\sigma}_b^{\dot{\beta}\beta} \left[[{\cal D}_{\beta}, 
\bar{{\cal D}}_{\dot{\beta}}] {\cal J}^a\right]_{\theta=0} =
\hspace{1.5cm} \nonumber \\ 
& = {1\over 64} 
\tilde{\sigma}_b^{\dot{\beta}\beta} \tilde{\sigma}_c^{ \dot{\alpha} {\alpha} }
 \left[[{\cal D}_{\beta}, 
\bar{{\cal D}}_{\dot{\beta}}] 
[{\cal D}_{{\alpha}}, \bar{{\cal D}}_{\dot{\alpha}}] {\cal K}^{ac}
\right]_{\theta=0} =
\nonumber \\ 
& = {1\over 64} 
\tilde{\sigma}_b^{\dot{\beta}\beta} \tilde{\sigma}_c^{ \dot{\alpha} {\alpha} }
 \left[[{\cal D}_{\beta}, 
\bar{{\cal D}}_{\dot{\beta}}] 
[{\cal D}_{{\alpha}}, \bar{{\cal D}}_{\dot{\alpha}}] 
(\theta)^2 (\bar{\theta})^2 
\right]_{\theta=0} \times \nonumber \\ 
& \qquad \times 
\int_{W^1}  
 l(\tau )[{1\over \hat{E}} \hat{E}^c_{\tau} \hat{E}^{a}
] \vert_{\hat{\theta}=0} 
\delta^4(x-\hat{x}) 
\;  .
\end{eqnarray}
In the WZ 
gauge (\ref{WZgauge}) 
(recall that it can be fixed simultaneously with the gauge 
(\ref{thGAUGE})), where Eqs. (\ref{WZgEff})--(\ref{WZ0gg}) as well as 
(\ref{Ber0=e}) and 
$E_{\underline{\alpha}}{}^{\breve{\beta}}\vert_{\hat{\theta}=0}= 
\delta_{\underline{\alpha}}{}^{\breve{\beta}}$  
are valid, Eq. (\ref{EiInt01}) reads
 \begin{eqnarray}\label{EiInt02} 
& e(x) 
R_{bc}{}^{ac} \vert_{\theta=0} = c 
\int_{W^1}  
 l(\tau )[ \hat{e}_{b\tau} \hat{e}^{a}] 
\delta^4(x-\hat{x}) \; , 
\end{eqnarray}
where $c$ is a constant and $ \hat{e}^{a}\equiv d\tau e_\tau^a
= d\hat{x}^\mu (\tau) e_\mu^a(\hat{x})$  is the pull--back 
to the worldline 
of the bosonic form $e^a=dx^\mu  e_\mu^a(x)= E^a \vert_{\theta =0}$. 
Eq. (\ref{EiInt02}) coincides with the one obtained from the 
 supergravity--bosonic particle coupled action  
provided by the sum of the component action for supergravity and the 
bosonic particle action \cite{BdAI1}, for the case 
$D=4$.

\section{Conclusions}

We have provided in this paper a fully dynamical description of the 
$D=4$, $N=1$ supergravity and the massless superparticle 
coupled system. It is given by the sum 
of the superfield supergravity action \cite{WZ78}  
and the Brink--Schwarz action \cite{BS81} for the massless superparticle. 
We have derived the complete set of superfield equations of motion 
for such a dynamical system.

The {\sl superfield} generalizations of the Rarita--Schwinger 
(gravitino) equation and of the Einstein equation both
acquire  source terms. These sources are determined by the 
Grassmann spinor covariant derivatives of one vector superfield 
${\cal J}_{a}$, the current `potential',
which is a current density distribution 
with support on the worldline that  appears at the right hand side 
of the vector superfield equation  
(\ref{4WGa}) for the supergravity--superparticle coupled system.

The current potential 
${\cal J}_{a}$ is covariantly conserved, Eqs. (\ref{DJa=0}), (\ref{DaJa=0}), 
and turns out to be constructed from the 
spin $3/2$ and spin $2$ distributions (\ref{K2}), (\ref{K3/2}), which we call 
`current prepotentials'.    
These current prepotentials obey
Eqs. (\ref{K2+}), (\ref{K3/2+}), (\ref{DK}),  as 
a result of the superparticle equations of motion. 

In the interacting system 
with dynamical supergravity, 
the Goldstone nature of the superparticle coordinate functions 
$\hat{Z}^M(\tau)$ \cite{Polchinski,Ivanov,Pashnev} 
allows one 
to fix the gauge (\ref{thGAUGE}) that sets the 
Grassmann 
coordinate function equal to zero, $\hat{\theta}(\tau)=0$ 
({\it cf.} \cite{Volkov}). 
The analysis of the local (gauge) symmetries of the coupled system 
shows that it is possible to fix simultaneously $\hat{\theta}(\tau)=0$ and 
the 
Wess--Zumino 
gauge for the supergravity variables. 
Clearly, with these gauge fixing conditions, 
after  integration over the superspace 
Grassmann coordinates $\theta$ 
and the elimination of the auxiliary fields $G_a\vert_{\theta=0}, 
R\vert_{\theta=0}, \bar{R}\vert_{\theta=0}$ by means of 
their (algebraic) equations of motion, 
the supergravity--superparticle interacting  action (\ref{SGSP}) 
should reduce to the action for supergravity--bosonic particle system 
investigated in \cite{BdAI1}. 
To verify this conclusion we have studied the component equations of motion 
derived from the superfield equations for 
the supergravity--superparticle system and shown that they 
do coincide with the supergravity bosonic particle 
equations from \cite{BdAI1} when both the WZ 
gauge
and the gauge (\ref{thGAUGE}) are used. In particular, 
in the resulting gauge the 
component Rarita--Schwinger equations remain sourceless 
while the Einstein equations acquire a source term from the 
(super)particle. 

The net outcome of our analysis is that 
the complete superfield action for the supergravity--superparticle
interacting system has the 
supergravity--bosonic particle system as its gauge fixed version, as
it is also the case for  
 the group--manifold based action for the coupled system \cite{BAIL}.

The applications of the present approach 
to the case of $D=4$ supergravity--superstring and
 supergravity--supermembrane systems 
requires a previous knowledge  of $D=4$ superspace supergravity 
with additional two--form and three--form in superspace
({\it cf.} \cite{G80}). This, as well as an analysis of the 
(super--)Higgs effect 
in the presence of superbranes 
and the study of the interaction of supergravity with more than one 
superbrane, will be the subject of future work.

\medskip 

{\it Acknowledgments}. 
The authors are grateful to D. Sorokin and  to 
I. Bars, E. Bergshoeff, M. Cederwall, G. Dvali, E. Ivanov, 
A. Pashnev, P. Pasti, K. Stelle and M. Tonin 
for useful discussions at different stages of this work. 
This research has been partially supported by 
the Spanish Ministry of Science and Technology through
grants BFM2002-03681, BFM2002-02000 and EU FEDER funds, 
the Junta de Castilla y Le\'on thorugh grant VA085-02, 
the Ucrainian FFR (research project $\# 383$), INTAS 
(research project N 2000-254) and by the KBN grant 5P03B05620.  
One of the authors (IB) thanks the Abdus Salam ICTP for their 
hospitality in Trieste during the final stages of this work.

\section*{Appendix A: Chiral projectors} 
\def\theequation{A.\arabic{equation}}
\setcounter{equation}0

The algebra of covariant derivatives ${\cal D}_A$, Eq. (\ref{4WD}), 
is encoded in 
the Ricci identities
\begin{eqnarray}
\label{RI}
& {\cal D}{\cal D}V_{A}= 
R_{A}^{\; B} \, V_{B} \; \leftrightarrow 
\cases{ {\cal D}{\cal D}V_{a}= 
R_{a}^{\; b} \, V_{b} \; , \cr 
{\cal D}{\cal D}V_{\alpha}= 
R_{\alpha}^{\; \beta} \, V_{\beta} \; , \cr 
{\cal D}{\cal D}V_{\dot{\alpha}}= 
R_{\dot{\alpha}}^{\; \dot{\beta}} \, V_{\dot{\beta}} \; , \cr } 
\end{eqnarray} 
where $V_A = (V_{a}, V_{\alpha},  V^{\dot{\alpha}})$ is an arbitrary 
supervector with 
tangent superspace Lorentz 
indices. 
Decomposing (\ref{RI}) on the basic two--forms $E^A \wedge E^B$, one finds 
(see \cite{1001,BW}) 
\begin{eqnarray}\label{algD} & 
 [ {\cal D}_{A}\; , {\cal D}_{B}\}V_{C} = - 
T_{{A}{B}}{}^{{D}} {{\cal D}}_{{D}}V_{{C}}+ 
R_{{A}{B}\;{C}}{}^{{D}}\, V_{{D}} \; . 
\end{eqnarray} 
When the constraints 
(\ref{4WTa=}), (\ref{4WTal=}),  (\ref{4WTdA=}), (\ref{4WR=}), 
(\ref{chR}), (\ref{chW}), (\ref{DG=DR}) are taken into account, 
Eqs. (\ref{RI}) (or   (\ref{algD}))   
 imply 
\begin{eqnarray}
\label{(DD)}
& \{ {\cal D}_\alpha , {\cal D}_\beta \} \; V_\gamma = - 
\bar{R} \epsilon_{\gamma (\alpha } V_{\beta)}\; , 
\\ 
\label{(DD)1}
& \{ {\cal D}_\alpha , {\cal D}_\beta \} \; V^\gamma = - 
\bar{R} V_{(\alpha}\delta_{\beta)}{}^{\gamma}\; , 
\\ 
\label{(DbD)} 
& \{ {\cal D}_\alpha , \bar{\cal D}_{\dot{\beta}}\}  = 
2i \sigma^a_{\alpha\dot{\beta}}{\cal D}_a \equiv 
2i {\cal D}_{\alpha\dot{\beta}}
\; ,  \qquad etc. 
\end{eqnarray}
In their turn, Eqs. (\ref{(DD)}), (\ref{(DD)1}) and their complex 
conjugates 
determine the form of the chiral projectors,  
{\it i.e.} they can be used to prove 
the identities
\begin{eqnarray}
\label{(DD)D}
& ({\cal D} {\cal D} - \bar{R}) \, 
{\cal D}_\alpha \xi^\alpha :=
({\cal D}^\beta {\cal D}_\beta - \bar{R}) \, 
{\cal D}_\alpha \xi^\alpha = 0\; , 
\\ \nonumber 
& (\bar{\cal D} \bar{\cal D} - {R}) \, 
\bar{\cal D}_{\dot{\alpha}} \bar{\xi}^{\dot{\alpha}} :=
(\bar{\cal D}_{\dot{\beta }} \bar{\cal D}^{\dot{\beta }} - {R}) \, 
\bar{\cal D}_{\dot{\alpha}} \bar{\xi}^{\dot{\alpha}} = 0\; , 
\\ \label{D(DD)}
& {\cal D}_\alpha ({\cal D} {\cal D} - \bar{R})U \, 
= 0\; , 
\\ \nonumber 
& \bar{\cal D}_{\dot{\alpha}} 
(\bar{\cal D} \bar{\cal D} - {R})U \, 
= 0\; ,  
\end{eqnarray}
where $\xi^\alpha, \bar{\xi}^{\dot{\alpha}}$ 
are arbitrary spinor superfields and 
$U$ is an arbitrary scalar superfield. 
Note that the chiral projectors are different when acting on superfields
with Lorentz group indices, {\it e.g.} 
\begin{eqnarray} \label{(DD)DU}
& ({\cal D} {\cal D} + {1\over 2}\bar{R}) \, {\cal D}_\alpha U \equiv  0\; .
\end{eqnarray}

\section*{APPENDIX B: 
On superdiffeomorphism invariance and  superspace general 
coordinate invariance}
\def\theequation{B.\arabic{equation}}
\setcounter{equation}0

In this Appendix we present a complete account of all the 
manifest superfield gauge symmetries of superfield supergravity.  
We discuss separately the  
{\sl active} and {\sl passive} forms of {\sl general coordinate invariance}  
which we call {\sl general coordinate symmetry}, $\delta_{gc}$, and 
{\sl superdiffeomorphism symmetry}, $\delta_{diff}$, respectively. 
Although both symmetries are known, usually only one of these two 
symmetries are considered in the literature.  
The reason is that the invariance 
of the Lagrangian form in a field theory (or of the 
Lagrangian integral form in a superfield theory) 
under $\delta_{diff}$  implies immediately  
the invariance under $\delta_{gc}$ (see Appendix A1 for further discussion). 
However, when dealing with a new type of 
system where some of the (super)fields live on a 
submanifold of (super)space ({\it e.g.}, on the  superparticle worldline) 
while others are defined on whole superspace, 
it is important to take into account that $\delta_{diff}$ and 
$\delta_{gc}$ act differently. 
In fact, this difference is already seen even for `free' supergravity 
where we show (Appendix B2) that the Wess--Zumino gauge is invariant under 
$\delta_{gc}$, 
whereas  the $\delta_{diff}$ transformations are broken by the 
Wess--Zumino gauge fixing conditions down to 
{\sl spacetime} local supersymmetry and {\sl spacetime}  
diffeomorphisms.

First, let us note that the set of superspace 
local Lorentz transformations
\begin{eqnarray}
\label{LorentzS}
&
\delta_{L} E^A= E^B L_B{}^A(Z) \quad \Leftrightarrow \hspace{2cm} \nonumber \\ 
& \cases{\delta_{L} E^a = E^b L_b{}^a(Z) \; , \; L^{ab}= - L^{ba}:=L^{ab}(Z) 
\cr 
\delta_{L} E^\alpha = E^\beta L_\beta{}^\alpha \; , \;  
L_\beta{}^\alpha = {1\over 4} L^{ab} 
(\sigma_a \tilde{\sigma}_b)_\beta{}^\alpha \; , 
\cr
\delta_{L} E^{\dot{\alpha}} = E^{\dot{\beta}} L_{\dot{\beta}}{}^
{\dot{\alpha}} \; , \; 
L_{\dot{\beta}}{}^{\dot{\alpha}}  = - {1\over 4} L^{ab} 
(\sigma_a \tilde{\sigma}_b)^{\dot{\alpha}}{}_{\dot{\beta}}\; , }
\nonumber \\ 
& \delta_{L} w^{ab}= {\cal D}L^{ab}\; , \hspace{2cm}  
\end{eqnarray} 
is a manifest symmetry of the constraints.  
Clearly,  they do not act on the superspace coordinates 
$\delta_{L} Z^M=0$.

Secondly, the constraints 
(\ref{4WTa=}), (\ref{4WTal=}),  (\ref{4WTdA=}), (\ref{4WR=}), 
(\ref{chR}), (\ref{chW}), (\ref{DG=DR}), 
as relations among differential forms,  are independent on the 
choice of a superspace local coordinate system. This evident statement 
can be formulated as an invariance under {\sl superdiffeomorphism} 
({\it i.e.},  superspace diffeomorphism) 
transformations $\delta_{diff}$ (see \cite{BdAI1}),    
\begin{eqnarray}
\label{sdZ+}
& Z^{\prime M}= Z^M + b^M(Z)\; : \quad 
\cases{
x^{\prime \mu}= x^\mu + b^\mu (x, \theta )\, , \cr 
\theta^{\prime \breve{\alpha}}=  \theta^{\breve{\alpha}} +  
\varepsilon^{\breve{\alpha}}
(x, \theta) \; , }\, 
\\ \label{sdiffF+}
& E^{\prime A}(Z^\prime) =  
E^{A}(Z), \quad  w^{\prime ab}(Z^\prime)=
w^{ab}(Z)\; .  \qquad 
\end{eqnarray} 
The statement of the invariance of differential forms, Eqs. (\ref{sdiffF}), 
\begin{eqnarray}
\label{sdiffZ+}
& \delta_{diff}Z^M=Z^{\prime M} - Z^M= b^M(Z)\; , \; 
\\ 
\label{sdiffF1}
& \delta_{diff} E^{A}=
E^{\prime A}(Z^\prime) - E^{A}(Z) = 0 \; , \quad 
\\ \label{sdiffF1w} & 
\delta_{diff} w^{ab} = w^{\prime ab}(Z^\prime)- w^{ab}(Z)=0 \; , \quad etc.,  
\end{eqnarray} 
just implies that Eq. (\ref{sdZ+}) (or (\ref{sdiffZ}))
describes a change of local coordinates, 
but does not act on 
the superspace `points' 
\footnote{The prime under differential form means, 
{\it e.g.}, $E^{\prime A} (Z^\prime) \equiv E^{A}(Z(Z^\prime))$. Thus Eq. 
(\ref{sdiffF1}) is the trivial identity 
$E^A(Z)\equiv E^{A}(Z(Z^\prime))$
reflecting the freedom of choosing an arbitrary  
local coordinate system. 
Nevertheless, the {\sl form invariance} of an action or of an equation 
under $\delta_{diff}$ requires the model to be formulated using the  
supervielbein superfield (in the bosonic case, when 
spinor fields are absent, 
it is enough to introduce a metric field). 
Thus  $\delta_{diff}$  can be used as a gauge 
principle for gravity and supergravity models.}.    
Thus $\delta_{diff}$ invariance can be treated as the {\sl passive} 
form of the general coordinate symmetry in superspace.

Thirdly, the set of constraints is invariant under  
{\sl general coordinate transformations of superspace} $\delta_{gc}$
\cite{WZ78,BW,BdAI1}
({\sl active} form of general coordinate symmetry). 
$\delta_{gc}$ is the symmetry 
under an arbitrary change of superspace `points' 
(in contrast to a change of local coordinates as 
in the case of $\delta_{diff}$)
\begin{eqnarray}\label{gcZ} 
& \delta_{gc} Z^M = & t^M(Z^M)\; . 
\end{eqnarray} 
The transformation of differential forms under the change of arguments 
(\ref{gcZ}) is given by the Lie derivative 
${\cal L}_t \equiv i_t d + d i_t$, {\it i.e.}, 
\begin{eqnarray}
\label{gcEA} 
& \delta_{gc} E^{A}(Z) := & E^{A} (Z+t) - E^{A}(Z)= 
{\cal L}_t E^{A}(Z) =
\nonumber \\ && =  i_t T^{A} + {\cal D}(i_t E^{A}) + 
E^{B} i_t w_{B}{}^{A} \; , 
\\ \label{gcwab} 
& \delta_{gc} w^{ab}(Z) := & w^{ab}(Z+t)- w^{ab}(Z)= {\cal L}_t w^{ab}(Z) = 
\nonumber \\ && = i_t R^{ab} +  {\cal D}(i_t w^{ab}(Z))\; , 
  \quad  etc. \, ,
\end{eqnarray}  
where 
\begin{eqnarray}\label{itEA}
& i_t E^A = t^M E_M^A =: t^A\; , \qquad {} \qquad \\ 
\label{itwab} & 
i_t w^{ab}(Z)=
t^M(Z) w_M^{ab}(Z)= t^A(Z) w_A^{ab}(Z)\; , \quad 
\\ \label{itTA}
& i_t T^A= E^B t^C T_{CB}{}^A \; , \quad i_t R^{ab}=E^B t^C R_{CB}^{ab}\; .
 \end{eqnarray}

The last terms in Eqs. (\ref{gcEA}), (\ref{gcwab}) 
 can be regarded as a Lorentz transformation (\ref{LorentzS})
induced by $\delta_{gc}$, 
$\delta_{L}(L^{ab}= i_t w^{ab})$ 
 and, thus, they can be conventionally  ignored in a manifestly Lorentz 
invariant theory. In other words, one may consider, equivalently, 
the superposition of transformations 
$\delta_{gc}(t) +\delta_{L}(L^{ab}= - i_t w^{ab})$ instead of 
the original $\delta_{gc}(t)$. 
These transformations were called 
{\sl supergauge transformations} in 
\cite{BW}. 

The simplest way to see that the constraints are invariant under 
the superspace general coordinate transformations 
$\delta_{gc}$ is to recall that $\delta_{gc}$ implies moving from one
superspace `point' to another one and that, since the constraints are 
satisfied 
at any superspace `point', 
they are invariant
\footnote{Denote the set of constraints 
by $C_2^{\cal A}(Z)\equiv {1\over 2} E^B \wedge E^C C_{CB}{}^{\cal A}(Z)=0$. 
They are satisfied at any superspace point $Z^M$. Thus,    
$C_2^{\cal A}(Z+t(Z))=0$ too and   $\delta_{gc} C_2^{\cal A}(Z)= 
C_2^{\cal A}(Z+t(Z))- C_2^{\cal A}(Z)=0$.}. 

Note also that the transformations of superforms, 
(\ref{gcEA}), (\ref{gcwab}), $  \delta_{gc}T^A= {\cal D}i_tT^A + 
i_t{\cal D}T^A$, {\it etc.}, imply the usual transformation rules for 
the (super)tensors (zero forms). For instance, for $T_{CB}{}^A$ defined by 
Eqs. (\ref{4WTa=def})--(\ref{4WTdA=def}),  
$T^A := {1\over 2} E^C \wedge E^B T_{BC}{}^A$, one obtains  
$ \delta_{gc}T_{CB}{}^A= t^D  {\cal D}_D T_{CB}{}^A$.

The fermionic general coordinate transformations (\ref{gcZ}),  (\ref{gcEA}),  
with parameter $t^M (Z) = \epsilon^{\underline{\alpha}} (Z) 
E_{\underline{\alpha}}{}^M (Z)$,  {\it i.e.} (see (\ref{itEA})),
\begin{eqnarray}\label{lsdef} 
& i_\epsilon E^a = 0 \; , \qquad i_\epsilon E^{\underline{\alpha}}= 
\epsilon^{\underline{\alpha}} (Z) \; ,
\end{eqnarray} 
can be treated as a  local supersymmetry \cite{BW}, while 
the bosonic transformations (\ref{gcZ}) 
with parameter $t^M (Z) = t^a (Z) 
E_{a}{}^M (Z) $ provide the superfield generalization of 
the spacetime general coordinate transformations. 
However, with such treatment, the origin of the local supersymmetry 
of the {\sl component} formulation of supergravity, {\it i.e.} of 
supergravity formulated as a theory of fields 
on spacetime, becomes slightly obscure. 
The following observation helps to 
make the above mentioned relation clearer.

Since 
diffeomorphism invariance $\delta_{diff}(b^M )$, 
(Eqs. (\ref{sdZ+}),  (\ref{sdiffF+})) is guaranteed, 
one can consider, instead of (\ref{gcZ}),  
the {\sl variational version of the  general coordinate transformations} 
\cite{WZ78} 
$\tilde{\delta}_{gc}$ with  parameter $t^A(Z) = t^ME_M^A:= i_t E^A$, 
defined by (see \cite{BdAI2})
\begin{eqnarray}\label{tgcdef}
\tilde{\delta}_{gc}(t^A) & = {\delta}_{gc}(t^M)
+ \delta_{diff}(b^M = -t^M )+ \nonumber \\ 
& \qquad {} \qquad + \delta_{L}(L^{ab}= - i_tw^{ab})\; .
\end{eqnarray}  
$\tilde{\delta}_{gc}(t)$ does not act on the superspace coordinates 
and acts on superforms 
through the covariant Lie derivative 
\begin{eqnarray}\label{tgcZ} 
& \tilde{\delta}_{gc} Z^M = & 0  \; , \\ 
\label{tgcE} 
& \tilde{\delta}_{gc} E^A(Z)  &=   i_t T^A + {\cal D}t^A \; , \\ 
\label{tgcwab}
& \tilde{\delta}_{gc} w^{ab}(Z)  &=   i_t R^{ab} \; , \quad  etc. \, .  
\end{eqnarray}

The {\sl superfield} local supersymmetry 
$\delta_{ls} (\epsilon^{\underline{\alpha}})$ can be identified with the 
variational version $\tilde{\delta}_{gc}(\epsilon^{\underline{\alpha}})$ 
of the fermionic general coordinate transformations 
(\ref{tgcZ}), (\ref{tgcE}): 
\begin{eqnarray}
\label{lsusydef}
\delta_{ls} (\epsilon^{\underline{\alpha}})= \tilde{\delta}_{gc}(t^a=0, 
t^{{\alpha}}=\epsilon^{{\alpha}}(Z), 
t^{\dot\alpha}= \bar{\epsilon}^{\dot{\alpha}}(Z) 
)\; .
\end{eqnarray}  
Then the relation with the local supersymmetry of the component formulation of 
supergravity becomes especially transparent.

Indeed, Eqs. (\ref{tgcE}), (\ref{tgcwab}) with the torsion and curvature 
 two--forms from 
 (\ref{4WTa=}), (\ref{4WTal=}), (\ref{4WTdA=}), (\ref{4WR=}), 
 and 
$t^A=(0, \epsilon^{\underline{\alpha}}(Z),\bar{\epsilon}^{\dot{\alpha}}(Z))$ 
provide us with the following local superspace supersymmetry 
transformations
\begin{eqnarray}
\label{lsusyx}
& \delta_{ls} Z^M =0\;  \qquad   \Leftrightarrow \qquad 
\cases{\delta_{ls} x^\mu =0\; , \cr
\delta_{ls} \theta^{\breve{\alpha}} = 0\; , }  
\\ \label{lsusyEa}
& 
\delta_{ls} 
 E^a = - 
2i E^{{\alpha}} 
\sigma^a_{{\alpha}\dot{\beta}} \bar{\epsilon}^{\dot{\beta}}(Z) -
2i \bar{E}^{\dot{\alpha}} 
\sigma^a_{\beta\dot{\alpha}} \epsilon^{{\beta}}
\; , \\  
\label{lsusyEal}
 & \delta_{ls} E^{{\alpha}} =
{\cal D} {\epsilon}^{{\alpha}} + 
{i\over 8} E^a [(\epsilon \sigma_a\tilde{\sigma}_b)^{{\alpha}} G^b 
+ (\bar{\epsilon}\tilde{\sigma}_a)^{\alpha} R]\; ,  \\  
\label{lsusyEbal}
 & \delta_{ls} \bar{E}^{\dot{\alpha}} = 
{\cal D} \bar{\epsilon}^{\dot{\alpha}} - 
{i\over 8} E^a [(\tilde{\sigma}_b\sigma_a \bar{\epsilon})^{\dot{\alpha}} 
G^b + ( \tilde{\sigma}_a\epsilon) ^{\dot{\alpha}} \bar{R}] \; , 
\end{eqnarray}
\begin{eqnarray}\label{lsusywff}
\delta_{ls} w^{\alpha\beta} & = - E^{(\alpha }\epsilon^{\beta )}\bar{R} - 
{i\over 8} E^a [(\tilde{\sigma}_a)^{\dot{\gamma}(\alpha } \epsilon^{\beta )}
\bar{{\cal D}}_{\dot{\gamma}}\bar{R} + \nonumber \\ 
& + (\epsilon \sigma_a\tilde{\sigma}_b)^{(\alpha} {\cal D}^{\beta )} G^b]\; .
\end{eqnarray}
The superspace local supersymmetry transformations 
$ \delta_{ls}$ of the main superfields 
(\ref{4WGa})--(\ref{4WchW}) are determined by 
\begin{eqnarray}
\label{lsusyR}
& \delta_{ls}R = \epsilon^{\alpha} {\cal D}_{\alpha}R \; , \quad 
 \delta_{ls}\bar{R}= 
\bar{\epsilon}^{\dot\alpha}\bar{{\cal D}}_{\dot\alpha}\bar{R} \; , 
\\ \label{lsusyGa}
& \delta_{ls}G^a = \epsilon^{\alpha} {\cal D}_{\alpha}G^a + 
\bar{\epsilon}^{\dot\alpha}\bar{{\cal D}}_{\dot\alpha}G^a \; ,  
\\ \label{lsusyW}
& \delta_{ls}W^{\alpha\beta\gamma}= \epsilon^{\delta} {\cal D}_{\delta}
W^{\alpha\beta\gamma}\; , \quad 
\delta_{ls}\bar{W}^{\dot{\alpha}\dot{\beta}\dot{\gamma}}= 
\bar{\epsilon}^{\dot\delta}\bar{{\cal D}}_{\dot\delta}
\bar{W}^{\dot{\alpha}\dot{\beta}\dot{\gamma}}\, .
\end{eqnarray}

Setting $\theta=0$ in the 
$ \delta_{ls}$ transformations (\ref{lsusyEa}) -- (\ref{lsusyW})
we arrive at the transformation rules of 
the {\sl off-shell} supersymmetry characteristic of the minimal 
formulation of the $D=4, N=1$ supergravity. To this end one needs 
the expression of the spinor derivatives of the main superfields in 
terms of the Riemann curvature ($R_{cd}{}^{ab}$, Eqs. (\ref{4WR=def}), 
(\ref{Rab})) and the gravitino field strengths ($T_{bc}^{\alpha},  
T_{bc}^{\dot\alpha}$, Eqs. (\ref{Tabg}), (\ref{Tabdg}))  
with the use of the consequences of the constraints 
(\ref{4WTa=}), (\ref{4WTal=}),  (\ref{4WTdA=}), (\ref{4WR=}), 
(\ref{chR}), (\ref{chW}), (\ref{DG=DR}). For instance, 
${\cal D}_{\alpha}{R}$ and  $\bar{{\cal D}}_{\dot\alpha}\bar{R}$ 
are expressed through the gravitino field strength 
$T_{ab\beta}$ with the use of Eqs. (\ref{gr1}).

The variational version of the 
superspace general coordinate transformations $\tilde{\delta}_{gc}$ 
with bosonic parameters $t^a$ can be called `local translations', 
${\delta}_{lt}= \tilde{\delta}_{gc}(t^a, t^\alpha =0)$ 
\begin{eqnarray}
\label{ltx}
& \delta_{lt} Z^M =0 \qquad \Leftrightarrow \qquad \cases{ 
\delta_{lt} x^\mu =0\; , \cr  
\delta_{lt} \theta^{\breve\alpha} = 0\; ,} \qquad 
\\ \label{ltEa}
& \delta_{lt} E^a  = {\cal D}t^a + 
{1\over 8} E_b \varepsilon^{abcd} t_c(Z) G_d \; , \qquad 
{} \qquad \\ \label{ltEal}  
& \delta_{lt} E^{{\alpha}}  = 
-  {i\over 8} E^{\beta} t^a (\sigma_a \tilde{\sigma}_b)_{\beta}
{}^{\alpha} G^b  + \qquad {} \qquad 
\nonumber \\ & \qquad 
{} \qquad + {i\over 8} \bar{E}^{\dot{\beta}} 
\epsilon^{\alpha\beta} t^a\sigma_{a\beta\dot{\beta}}R 
+ E^b t^a T_{ab}{}^{\alpha}\; , \\ \nonumber & etc. 
\end{eqnarray}
In the pure bosonic case it is precisely this symmetry 
(this form of the spacetime general coordinate invariance) 
that provides the possibility of treating general relativity 
as a gauge theory of the 
Poincar\'e group \cite{GRgauge} (see \cite{BdAI2} for further discussion).

\subsection*{B1. Gauge symmetries of the `free' supergravity 
superfield action}

The action (\ref{SGact}) is evidently invariant under 
the superdiffeomorphisms
(\ref{sdZ+}), (\ref{sdiffF+}), 
\begin{eqnarray}\label{sdiffSSG}
\delta_{diff} S_{SG} = 0 \; . 
\end{eqnarray} 
This invariance is a simple consequence of the possibility of changing 
variables in {\sl any} integral (see footnote 11); but moreover, 
in our case the action is {\it form invariant} as the theory is 
formulated in terms of the supervielbein.

In the pure bosonic case, where the counterpart of the above statement means 
that the action is an integral of a differential form (Lagrangian form), 
$S= \int_{M^D} L_{D}$, the general coordinate invariance follows then 
from the simple observation that the variation of the Lagrangian form under 
$\delta_{gc}$, as well as under $\tilde{\delta}_{gc}$, is given 
(see \cite{BdAI2}) by a Lie derivative:
$\delta_{gc}L_D \equiv  \tilde{\delta}_{gc}L_D = i_t dL_D  + 
d(i_t L_D )$. Then the first term vanishes as it contains 
the exterior derivative 
of $D$--form on a $D$--dimensional manifold, while the second term 
is a total derivative which does not contribute 
 for a spacetime 
$M^D$ without boundary.
This statement is usually treated 
as a manifestation of the equivalence between the active and passive 
forms of general 
coordinate transformations. However, 
although these symmetries imply each  
other in  field theories, their  
r\^ole is different as we show in Appendix C (C2, C3).

In the case of superspace the action is written in terms of an  
integral (Berezin) form.  
Nevertheless, the general coordinate invariance of the 
superdiffeomorphism invariant action can be also established easily. 
For instance, to prove the invariance of the action (\ref{SGact}) 
under the variational version 
$\tilde{\delta}_{gc}(t^A)$ 
(\ref{tgcZ}), (\ref{tgcE}), (\ref{tgcwab})
of the superspace general coordinate 
transformations, including local   
supersymmetry (Eqs. (\ref{lsusyx}), (\ref{lsusyEa})--(\ref{lsusyW})),   
one has to use the identity 
\begin{eqnarray}
\label{idd80}
\int d^4x \tilde{d}^4 {\theta}    
E \; ({\cal D}_{A}\xi^{A} + \xi^B T_{BA}{}^A) (-1)^{A}\; \equiv 0 \; ,
\end{eqnarray}
which is valid for any complex superfield $\xi^{A}=(C^a(Z), 
\nu^\alpha (Z), \bar{\mu}^{\dot{\alpha}}(Z))$. 

A variation of superdeterminant 
has the form
\begin{eqnarray}\label{vBer}
\delta E \equiv \; E \; E_A{}^M \, \delta E_M{}^A\; (-1)^A
\; . 
\end{eqnarray}  
To compute $\tilde{\delta}_{gc} E$ one 
ubstites  $i_M(\tilde{\delta}_{gc}E^A)$ from 
Eqs. 
(\ref{tgcE}) for $\delta E_M{}^A$, and finds 
\begin{eqnarray}\label{gcBer}
\tilde{\delta}_{gc} E = E (-1)^A {\cal D}_A t^A  + 
E (-1)^A t^B T_{BA}{}^A \; . 
\end{eqnarray}  
Then the identity (\ref{idd80}) implies 
\begin{eqnarray}\label{vgcSSG=0}
\tilde{\delta}_{gc} S_{SG}=
\int d^8Z   
E \; ({\cal D}_{A}t^{A} + t^B T_{BA}{}^A) (-1)^{A}= 0 \; .
\end{eqnarray}
This completes the proof of general coordinate 
symmetry.   

Note that, as the constraints of minimal supergravity 
(\ref{4WTa=}), (\ref{4WTal=}) imply 
$(-1)^A T_{BA}{}^A=0$, the identity (\ref{idd80}) simplifies to 
\begin{eqnarray}
\label{idd8}
\int d^8Z   
E \; {\cal D}_{A}\xi^{A}(-1)^{A}\; = 0 \; . 
\end{eqnarray}

Since $\delta_{ls}= \tilde{\delta}_{gc}(t^A = 
(0, {\epsilon}^{\underline{\alpha}}))$, 
Eq. (\ref{lsusydef}), 
this proves, in particular, the invariance under the local supersymmetry 
transformations (\ref{lsusyx})--(\ref{lsusyW}) 
(which imply, {\it e.g.},  
$\delta_{ls} 
 E_M^a = - 
2i E_M^{{\alpha}} 
\sigma^a_{{\alpha}\dot{\beta}} \bar{\epsilon}^{\dot{\beta}}(Z) + 
2i \bar{E}_M^{\dot{\alpha}} 
\sigma^a_{\beta\dot{\alpha}} \epsilon^{{\beta}}$). 
Specifically, one finds 
\begin{eqnarray}\label{lsSGact}
& \delta_{ls} S_{SG} = - \int d^8Z   
E \; {\cal D}_{\underline{\alpha}}{\epsilon}^{\underline{\alpha}} = 
\int d^8Z  {\cal D}_M (EE_{\underline{\alpha}}^M) \; 
{\epsilon}^{\underline{\alpha}} \nonumber  \\ 
&  \qquad \equiv  
- \int d^8Z T_{\underline{\alpha}{A}}{}^{A} (-1)^{A} 
{\epsilon}^{\underline{\alpha}} = 0 \; .
\end{eqnarray}

\subsection*{B2. On the gauge symmetries of the 
supergravity--superparticle coupled system}

The invariance of the coupled action under superspace diffeomorphisms 
$\delta_{diff}$ follows from the fact that 
Eqs. (\ref{sdiffhZ}), (\ref{sdiffZ}), (\ref{sdiffF1}) imply 
\begin{eqnarray}\label{sdiffhEa}
\delta_{diff} \hat{E}^{a} =  
\hat{E}^{\prime a}(\hat{Z}+ \delta_{diff} \hat{Z})- \hat{E}(\hat{Z})=0 \; .
\end{eqnarray}
Thus, 
\begin{eqnarray}\label{sdiffS0}
& \delta_{diff} S_{sp} = 0 \;  
\end{eqnarray}
and, since $\delta_{diff} S_{SG} = 0$ (Eq. (\ref{sdiffSSG})) 
we find 
\begin{eqnarray}\label{sdiffS}
& \delta_{diff} S = 0 \; . 
\end{eqnarray}

On the other hand, as the superspace coordinates $Z^M$ 
(not to be confused with  $\hat{Z}^M(\tau)$) 
do not enter in the superparticle action, the general coordinate 
transformations 
$\delta_{gc}$  (Eqs. (\ref{gcZ}), (\ref{gcEA})) 
 supplemented by the definition
\begin{eqnarray}\label{dgchZM}
& \delta_{gc} \hat{Z}^M(\tau) = 0 \; , 
\end{eqnarray}
trivially give ${\delta}_{gc} S_{sp}=0$, and 
the invariance of the supergravity action 
${\delta}_{gc} S_{SG}=0$ gives  
\begin{eqnarray}\label{vgcS+=0}
{\delta}_{gc} S =0\; . 
\end{eqnarray}
Then the invariance under the variational copy of the 
superspace general coordinate transformations, $\tilde{\delta}_{gc}$, 
Eqs. (\ref{tgcZ}), (\ref{tgcE}) supplemented by 
the definition
\begin{eqnarray}\label{tgchZ=2} 
& \tilde{\delta}_{gc} \hat{Z}^M(\tau)  = - t^M(\hat{Z}) 
\hspace{2cm} \\ \nonumber 
& \equiv 
 -t^a(\hat{Z})E_a^M(\hat{Z}) 
- \epsilon^{{\alpha}}(\hat{Z}) E_{{\alpha}}^M(\hat{Z}) 
-  \bar{\epsilon}{}^{\dot{\alpha}}(\hat{Z})
E_{\dot{\alpha}}^M(\hat{Z})\; , 
 \end{eqnarray}
follows from the $\delta_{gc}$ and $\delta_{diff}$ invariances 
\footnote{The breaking of $\tilde{{\delta}}_{gc}$ invariance,     
discussed in \cite{BdAI1,BdAI2}, is a spontaneous symmetry breaking.}, 
\begin{eqnarray}\label{vtgcS+=0}
\tilde{{\delta}}_{gc} S\equiv \tilde{\delta}_{gc} S_{sp}=0\; . 
\end{eqnarray}
Note that 
in the `superparticle sector' of the 
configuration space of the interacting system the 
action of $\tilde{{\delta}}_{gc}$, Eq. (\ref{tgchZ=2}),   
coincides with the action of diffeomorphism transformations. 

In particular, the transformation  
of the fermionic coordinate function 
$ \hat{\theta}^{\breve\alpha}(\tau)$ 
under the full set of 
local symmetries of the interacting system 
(including $\delta_{diff}(b)$ (\ref{sdiffhZ}), 
$\tilde{\delta}_{gc}(t)$ (\ref{tgchZ=2})
and the worldline $\kappa$--symmetry) 
acquires the form 
\begin{eqnarray}\label{thG0+}
\delta  \hat{\theta}^{\breve\alpha}(\tau)=  
b^{\breve\alpha}(\hat{Z}) -  t^{\breve\alpha}(\hat{Z}) 
+ \delta_{\kappa}  \hat{\theta}^{\breve\alpha}(\tau)
\; , 
\end{eqnarray} 
where $\delta_{\kappa}  \hat{\theta}^{\breve\alpha}(\tau)$ is defined by the 
Eq. (\ref{4kappa}) with 
$M= \breve{\alpha}$.

\section*{APPENDIX C: More on the Wess--Zumino gauge}  

\subsection*{C1. Decomposition of superfields in the Wess--Zumino gauge}
\def\theequation{C.\arabic{equation}}
\setcounter{equation}0 

The decomposition of the superfields 
$E_{\breve{\alpha}}{}^a$, 
$E_{\breve{\alpha}}{}^{\underline{\alpha}}$, $w_{\breve{\alpha}}{}^{ab}$
in power series on $\theta$ 
is completely determined  by 
Eqs. (\ref{WZg1b}), (\ref{WZg1f}), (\ref{WZg1w}). 
To make such an expansion explicit one can 
use the formal operator \cite{BZ85} 
\begin{eqnarray}\label{WZg(1+)}
& {1\over (1+\theta\partial)} = 
{1\over (1+\theta^{\underline{\alpha}}{\cal D}_{\underline{\alpha}})}\; 
\end{eqnarray}
The action of such an operator 
is well defined on superfields (as
they are  
polynomials in $\theta$) 
and produces  expressions involving {\sl covariant} Grassmann 
derivatives ${\cal D}_{\underline{\alpha}}$ when (\ref{WZg(1+)}) 
acts on 
the torsion and curvature superfields. For instance, from 
Eq. (\ref{WZg1b}) one finds  
 \begin{eqnarray} 
\label{WZgcb} 
& E_{\breve{\alpha}}^a (Z) = {1\over (1+\theta\partial)} 
\theta^{\underline{\beta}} T_{\underline{\beta}\breve{\alpha}}^a 
= \theta^{\underline{\beta}} 
{1\over (2+\theta^{\underline{\alpha}}{\cal D}_{\underline{\alpha}})}
(E_{\breve{\alpha}}^C T_{C\underline{\beta}}^a )\; , 
\end{eqnarray}
where we use 
Eq. (\ref{WZgD}) and the identity
 \begin{eqnarray}\label{WZid0} 
& {1\over (k+\theta\partial)} \theta^{\underline{\beta}}
\equiv \theta^{\underline{\beta}}
{1\over (k+1+\theta\partial)} \;  
\end{eqnarray}
(which follows from $(k+\theta\partial) \theta^{\underline{\beta}}= 
\theta^{\underline{\beta}} (k+1+\theta\partial)$).

As one more example, let us 
present the explicit form of the $d\theta$ component of the expression 
(\ref{WZgDth0}), which enters (\ref{WZg1f}): 
 \begin{eqnarray}
\label{WZgDth}
& {\cal D}_{\breve\alpha} \theta^{\underline{\beta}} =   
\delta_{\breve\alpha}^{\underline{\beta}}+ 
\theta^{\underline{\gamma}} w_{\breve{\alpha}\underline{\gamma}}
{}^{\underline{\beta}}
=  \qquad \nonumber \\ & 
\qquad {} \qquad  = \delta_{\breve\alpha}^{\underline{\alpha}}
(\delta_{\underline{\alpha}}^{\underline{\beta}}+ {1\over 4} 
\theta^{\underline{\gamma}}\Gamma_{ab\underline{\gamma}}{}^{\underline{\beta}}
{1\over (1+\theta \partial )} \theta^{\underline{\beta}} 
R_{\underline{\beta} \underline{\alpha}}{}^{ab})\;  
\nonumber \\ & 
\qquad {} \qquad  = \delta_{\breve\alpha}^{\underline{\alpha}}
(\delta_{\underline{\alpha}}^{\underline{\beta}}+  
{1\over 4} 
\theta^{\underline{\gamma}}\Gamma_{ab\underline{\gamma}}{}^{\underline{\beta}}
\theta^{\underline{\beta}} 
{1\over (2+\theta ^{\underline{\epsilon}}{\cal D}_{\underline{\epsilon}} )} 
R_{\underline{\beta} \underline{\alpha}}{}^{ab})\; . 
\end{eqnarray}

The complete decomposition of the $dx^\mu$ components of 
the forms  
$E^a$, $E^{\underline{\alpha}}$, $w^{ab}$ 
is governed by the $dx^\mu$ components of 
the 
(\ref{WZg1b}), (\ref{WZg1f}), (\ref{WZg1w}), 
{\it e.g.} 
\begin{eqnarray} 
\label{WZg2b} 
& \theta\partial E_\mu^a = E^B_\mu \theta^{\underline{\beta}} 
T_{\underline{\beta}B}{}^a = - 
\theta^{\underline{\beta}} E^B_\mu 
T_{B\underline{\beta}}{}^a
\; .
\end{eqnarray}
Clearly, Eq. (\ref{WZg2b}) involves  
the nilpotent operator $\theta\partial \equiv
\theta^{\breve{\alpha}}\partial_{\breve{\alpha}}= 
\theta^{\underline{\alpha}}{\cal D}_{\underline{\alpha}}$.  
This nilpotent operator has an evident kernel: the leading component of the 
superfield, {\it e.g.}  $E_m^a \vert_{\theta=0}=e_m^a(x)$. 
However, as it was observed in \cite{BZ85}, 
this operator can be considered as {\sl invertible in the 
 space of 
superfields with the vanishing leading components}. Thus 
one can write as well the formal expansion for 
$E_\mu^a$ 
by subtracting the kernel,  
$E_\mu^a(Z) \rightarrow (E_\mu^a(Z)- E_\mu^a \vert_{\theta=0})$
(thus arriving at a superfield with vanishing leading component)
and using the formal relation (\ref{WZid0}) with 
$k=0$ (which is meaningful in the  space of 
superfields with vanishing leading components) 
to arrive at
\begin{eqnarray} 
\label{WZg3b} 
& E_\mu^a(Z) = E_\mu^a \vert_{\theta=0} - 
\theta^{\underline{\beta}}
{1\over (1+\theta^{\underline{\alpha}}{\cal D}_{\underline{\alpha}})} 
(E^B_\mu (Z) T_{B\underline{\beta}}{}^a) \; .
\end{eqnarray}

\subsection*{C2. Gauge symmetries preserving the Wess--Zumino gauge}

To find the full set 
of  local symmetries that preserve the 
WZ gauge (\ref{WZgauge})
\footnote{Note that we do not use here the `prepotentail' 
form of the WZ gauge described in footnote 4, 
and shall not address the issues of residual symmetries in such gauge. 
This requires a separate study as a number of gauge symmetries have to be 
fixed before one arrives at the expression in terms of auxiliary 
vector superfield and chiral compensator, and,  on the other hand, 
the solutions of the constraints are defined modulo additional 
gauge symmetry 
transformations. Thus 
all our statements below are for
the Wess--Zumino gauge (\ref{WZgauge}) fixed through the conditions 
on the {\sl potentials} 
of the superfield supergravity.}
one may write the infinitesimal variations 
$\delta_{diff}(b^M)$,  
$\tilde{\delta}_{gc}(t^A)$,  
$\delta_{L}(L^{ab})$
of the WZ 
conditions (\ref{WZgauge}) and require their preservation, 
 \begin{eqnarray}\label{WZgpresE}
& (\theta^{\breve{\alpha}} + b^{\breve{\alpha}}(Z)) \, 
(E^{\prime A}_{\breve{\alpha}}(Z^\prime) + 
\delta_{L} E_{\breve{\alpha}}^A(Z) +\tilde{\delta}_{gc} 
E_{\breve{\alpha}} ^A(Z))=  \hspace{-1.5cm} \nonumber \\
& \qquad = (\theta^{\breve{\alpha}} + b^{\breve{\alpha}}(Z)) \, 
\delta_{\breve{\alpha}}^A \; , 
\\ \label{WZgpresw}
& (\theta^{\breve{\alpha}} + b^{\breve{\alpha}}(Z)) \, 
(w^{\prime ab}_{\breve{\alpha}}(Z^\prime) + 
\tilde{\delta}_{gc}w^{ab}_{\breve{\alpha}}(Z) + 
 \nonumber \\ 
& \hspace{4.5cm} + 
\delta_{L} w^{ab}_{\breve{\alpha}}(Z))=0\; .
\end{eqnarray}
Here $\tilde{\delta}_{gc}$ is defined by 
Eqs. (\ref{tgcZ}),  (\ref{tgcE}), (\ref{tgcwab})   
and  the primes reflect 
the superdiffeomorphism transformations, 
Eqs. (\ref{tgcZ}), (\ref{sdiffF}) or  (\ref{sdiffF1}), 
(\ref{sdiffZ}). Hence  
\begin{eqnarray}
 \label{sdiffFE}
& E_{\breve{\alpha}}^{\prime A}(Z^\prime) =
 E_{\breve{\alpha}}^{A}(Z) - 
\partial_{\breve{\alpha}} b^M \, 
E_{M}^{A}(Z) 
\; , 
\\
 \label{sdiffFw}
& w_{\breve{\alpha}}^{\prime ab}(Z^\prime)= 
w_{\breve{\alpha}}^{ab}(Z)
 - \partial_{\breve{\alpha}} b^M \, 
w_{M}^{ab}(Z) 
\; .
\end{eqnarray} 
The terms 
$\tilde{\delta}_{gc}E^{A}_{\breve{\alpha}}(Z)$ and 
$\tilde{\delta}_{gc}w^{ab}_{\breve{\alpha}}(Z)$ in Eq. 
(\ref{WZgpresE}), (\ref{WZgpresw})
are defined by the contraction 
of Eqs. (\ref{tgcE}), (\ref{tgcwab}),    
\begin{eqnarray}
\label{tgcEf} 
& \tilde{\delta}_{gc} E_{\breve{\alpha}}^A(Z)  =
 t^{B}T_{B\breve{\alpha}}^A + 
 {\cal D}_{\breve{\alpha}} t^A \; ,
\\ 
\label{tgcwabf}
& \tilde{\delta}_{gc} w_{\breve{\alpha}}^{ab}(Z)  =  
 t^D R_{D \breve{\alpha}}^{ab} \; .   
\end{eqnarray}  
Finally, the Lorentz transformations have the standard form (\ref{LorentzS}), 
(\ref{Ldec}), 
\begin{eqnarray}\label{LorE}
& \delta_{L} E_{\breve{\alpha}}^A(Z)= 
E_{\breve{\alpha}}^B(Z)L_B{}^A (Z) \; ,  
\\ \label{Lorw} & \delta_{L}w^{ab}_{\breve{\alpha}}(Z) = 
{\cal D}_{\breve{\alpha}}L^{ab}(Z)\; , 
\\ \label{Lordec}
& L_B{}^A (Z)= \left(\matrix{L_b{}^a & 0 \cr 
              0 & L_{\underline{\beta}}{}^{\underline{\alpha}}}\right)\; ,
\quad L^{ab}= - L^{ba}\; , \nonumber \\ & 
L_{\underline{\beta}}{}^{\underline{\alpha}}= 
{1\over 4} L^{ab} \gamma_{ab}{}_{\underline{\beta}}{}^{\underline{\alpha}}\; .
\end{eqnarray}

By algebraic manipulation with the use of 
the recurrent relations (\ref{WZg1b}), (\ref{WZg1f}), (\ref{WZg1w}), 
one can present Eqs. (\ref{WZgpresE}), (\ref{WZgpresw}) in the form 
\begin{eqnarray}
\label{WZgpr1+} 
& \theta \partial(b^A - t^A) = (b^B - t^B) 
\theta^{\underline{\gamma}} T_{\underline{\gamma}B}{}^A 
+ \qquad \nonumber \\ & \qquad 
+ \theta^{\underline{\gamma}} (L_{\underline{\gamma}}^{\underline{\beta}}
- b^M w_{M\underline{\gamma}}{}^{\underline{\beta}}) \; , 
\\ \label{WZgpr2+} 
&  \theta \partial(L^{ab}(Z)- b^M w_M{}^{ab})=  - 
(b^D-t^D) \theta^{\underline{\gamma}}  
R_{C\underline{\gamma}D}{}^{ab} \; .
\end{eqnarray}
Setting $t^A=0$ 
in Eqs. (\ref{WZgpr1+}), (\ref{WZgpr2+}), one arrives at 
Eqs. (\ref{WZgpr1}) (\ref{WZgpr2}).

At zero--order of the weak field approximation 
one finds the set of equations  ({\it cf.} (\ref{WZgrs1})--(\ref{WZgrs3})) 
\begin{eqnarray}
\label{WZgrs1+} 
& \theta \partial(b^a - t^a) = - 2i 
(\varepsilon^{\underline{\beta}} - \epsilon^{\underline{\beta}})
\gamma^a_{\underline{\beta}\underline{\gamma}}
\theta^{\underline{\gamma}} \; , 
\\ \label{WZgrs2+} 
&  \theta \partial(\varepsilon^{\underline{\alpha}}- 
\epsilon^{\underline{\alpha}}) = 
\theta^{\underline{\beta}} L_{\underline{\beta}}{}^{\underline{\alpha}}\; , 
\\ \label{WZgrs3+} 
&  \theta \partial L^{ab}(Z)= 0\; , 
\end{eqnarray}
which are solved by 
\begin{eqnarray}
\label{WZgrs1s+} 
t^a(Z) - b^a(Z) & = t_-^a(x) - 2i \theta\gamma^a\epsilon_-(x) - 
\hspace{1cm} \nonumber 
\\ & -
{i \over 4} \theta (\gamma_{bc}\gamma^a)\theta \, l^{bc}(x) \; , 
\hspace{1cm} \\ \label{WZgrs2s+} 
\epsilon^{\underline{\alpha}}(Z) - \varepsilon^{\underline{\alpha}}(Z)
& = 
\epsilon_-^{\underline{\alpha}}(x) - \theta^{\underline{\beta}}
l_{\underline{\beta}}{}^{\underline{\alpha}}(x)  \; ,
\hspace{1.3cm}
\\ \label{WZgrs3s+} 
L^{ab}(Z) & = l^{ab}(x)\; ,  \hspace{3.3cm}
\end{eqnarray}
where $t_-^a(x)$, $\epsilon_-^{\underline{\alpha}}(x)$ are arbitrary 
vector and spinor functions and $l^{ab}(x)$ are local Lorentz parameters.

In the general case the WZ 
gauge is preserved by the part of the original superspace local symmetry 
corresponding  to the parameters 
that are not restricted by Eqs. (\ref{WZgpr1}),  (\ref{WZgpr2}). 
These are the {\sl superfield} 
parameter  
\begin{eqnarray}\label{parS+}
t_+^A(Z)= b^A(Z) + t^A(Z) \; , 
\end{eqnarray} 
the vector and spinor {\sl field} parameters  
\begin{eqnarray}\label{parc+}
t_-^A(x)= (t_-^a (x), \epsilon_-^{\underline{\alpha}}(x))= 
(t^A(Z) - b^A(Z))\vert_{\theta=0} \; ,
\end{eqnarray}  
and the antisymmetric tensor  {\sl field} parameter
\begin{eqnarray}
\label{parL}
l^{ab}(x)= L^{ab}(Z)\vert_{\theta=0}\; .
\end{eqnarray} 

In particular, both the spacetime diffeomorphisms 
and general coordinate transformations (with parameters 
$b^a(Z)\vert_{\theta=0}$, $t^a(Z)\vert_{\theta=0}$), as well as 
Lorentz ($l^{ab}(x)$) and local supersymmetry 
($\epsilon^{\underline{\alpha}}(x)= \epsilon^{\underline{\alpha}}(Z)
\vert_{\theta=0}$) transformations preserve the WZ  
gauge.

\subsection*{C3: On the 
general coordinate invariance of the Wess--Zumino gauge}

The fact that the conditions (\ref{WZgpr1}), (\ref{WZgpr2}) 
on the parameters of the symmetry
that preserve the WZ gauge 
do not restrict also the superfield parameter (\ref{parS+}) 
requires some comments. 
Eq. (\ref{tgcdef})  
can be rewritten as
\begin{eqnarray}\label{tgcdef+}
{\delta}_{gc}(t^M) & = \tilde{\delta}_{gc}(t^A) 
+ 
\delta_{diff}(b^M = t^M )+ 
\nonumber \\ 
& \qquad {} \qquad + 
\delta_{L}(L^{ab}= i_tw^{ab})\; .
\end{eqnarray} 
Then, Eqs.  (\ref{WZgpr1}),  (\ref{WZgpr2}) become the identity 
$0\equiv 0$ when 
$b^M = t^M$, $L^{ab}= i_tw^{ab}$. 
This observation shows 
that the superfield symmetry which 
preserves the WZ 
gauge is just the general coordinate symmetry 
in its original form ${\delta}_{gc}$, Eqs. (\ref{gcZ}), (\ref{gcEA}), 
(\ref{gcwab}). 
This can be also verified straightforwardly.
   
Actually, the general coordinate invariance of the WZ 
gauge 
(\ref{WZgauge}) 
is natural and should be expected if one has in mind that 
the general coordinate transformations imply passing from one `point' of 
superspace to another, while the WZ 
gauge (\ref{WZgauge}) 
is valid at any superspace  `point'\footnote{
A bosonic counterpart of the above statement is that the defining conditions 
of the normal coordinate system in 
general relativity, $x^\mu (e_\mu^a(x)-\delta_\mu^a)=0$,  
$x^\mu \omega_\mu^{ab}=0$, are invariant under 
the active form of spacetime general coordinate transformations 
$x^\mu \rightarrow x^\mu + t^\mu(x)\,$, 
$\; e^a(x):= dx^\mu e_\mu^a(x)  \rightarrow e^a(x+t)= e^a(x) + i_tde^a + 
di_te^a$. This again can be easily explained by observing that the above 
conditions are valid at any spacetime point and that 
the active form of the general coordinate transformation implies just 
replacement of one spacetime point by 
another.}. 
 
It is instructive to understand how this symmetry is realized in the 
spacetime  supergravity action. Let us consider first a superfield 
action with a full superspace (Berezin) measure ({\it e.g.} the 
functional (\ref{SGact})) which possesses superspace 
general coordinate invariance. Then one can integrate over the  
Grassmann variables and arrive at a component action 
written as the integral over spacetime of a Lagrangian form 
expressed in terms of 
spacetime fields. However, as this is still {\sl the same} action, it should 
still possess the {\sl superspace} general coordinate invariance. 
But, on the other hand, it is independent 
of the Grassmann variables after the Berezin integration. 
The resolution of  this apparent paradox is that on 
the component fields 
the {\sl superspace} general coordinate 
transformations (\ref{gcZ})--(\ref{gcEA})  are realized nonlinearly, 
with only the subgroup 
of {\sl spacetime} general 
coordinate transformations acting linearly. 
For instance, on the spacetime vielbein form $e^a(x)= 
E^a(Z)\vert_{\theta =0, d\theta =0}$ the superspace general coordinate 
symmetry with parameters $t^M(Z)= (t^\mu (x, \theta), \epsilon^{\check{\alpha}}(x, \theta))$ acts as 
$e^a(x)\rightarrow 
e^a (x+t(x,\theta))\vert_{\theta + \epsilon(x,\theta)=0}$ 
({\it  cf.} the nonlinear realization of the superspace supergravity 
supergroups in \cite{IK};  it is instructive to note that 
the above expression simplifies if the superfield 
$\epsilon^{\check{\alpha}}$ is assumed to be independent of $\theta$, 
$\epsilon^{\check{\alpha}}(x, \theta)
= \epsilon^{\check{\alpha}}_0(x)$; in this case one finds 
$e^a(x)\rightarrow e^a (x+t(x,- \epsilon_0(x))$).

The r\^ole of superdiffeomorphism symmetry is different. It allows us  
to choose a coordinate  system in superspace 
(the WZ gauge) where all the higher terms in the decomposition 
of supervielbein superfields on powers of Grassmann coordinates are expressed 
in terms of leading components of 
supertensors (torsion, curvature and their covariant derivatives). 
 
The additional hidden  superspace general coordinate invariance of the 
component supergravity action may shed some light on the 
transition from the superfield action 
to its component form that uses   `Ectoplasm' ideas 
 \cite{NormalC1,SiBook}, as well as 
on the existence of the rheonomic or group manifold 
approach to supergravity \cite{rheo}\footnote{
The rheonomic approach to 
supergravity \cite{rheo} is based on a generalized 
action principle constructed in accordance with the following prescriptions  
(see \cite{bsv,BAIL}):  
i) one takes the usual component action, ii) writes it in the first order 
form, without using the Hodge duality operator, and 
iii) replaces all the fields 
by superfields, but taken on the surface 
${\cal M}^D$ 
in $D$--dimensional superspace 
defined paramerically by $\theta = \tilde{\theta}(x)$, 
where  $\tilde{\theta}(x)$ 
is an arbitrary fermionic function of spacetime 
coordinates. 
Such an action is clearly invariant under superspace general 
coordinate transformations pulled back onto the surface  ${\cal M}^D$. 
On the other hand, setting $\tilde{\theta}(x)=0$ one arrives at the 
first order form of the component action, where the superspace 
general coordinate invariance is not manifest, but it is a hidden symmetry 
allowing to go back to an arbitrary surface ${\cal M}^D$ in superspace 
({\it cf.} the {\sl rheonomic principle} of \cite{rheo}).} 
and of a related treatment of the $D=10$ superfield superstring action
\cite{Tonin}.

\section*{APPENDIX D: 
On worldline symmetries of the Brink--Schwarz   
superparticle action}
\def\theequation{D.\arabic{equation}}
\setcounter{equation}0

The reparametrization symmetry $\delta_{r}$, Eqs. (\ref{rep}), 
(\ref{repl}), is  the gauge symmetry of the superparticle 
action which can be identified with the 
{\sl variational version of the worldline general coordinate 
transformations},  $\tilde{\delta}_{wgc}$, 
because the transformations (\ref{4rep}), (\ref{rep}), (\ref{repl}) do not 
act on the proper time $\tau$. 
Note that, actually, 
as the natural definition of 
$\tilde{\delta}_{wgc}(s(\tau))$ ({\it cf.}  
(\ref{tgcZ}), (\ref{tgcE}))  is provided by 
\begin{eqnarray}\label{wlgct} 
\tilde{\delta}_{wgc} \tau := 0 \; , 
\\ \label{wgcZ} 
\tilde{\delta}_{wgc} \hat{Z}^M(\tau) = s(\tau)\partial_\tau\hat{Z}^M(\tau)
\; , \\ 
\label{wlgcl} 
\tilde{\delta}_{wgc}l(\tau)= l \partial_\tau s - s \partial_\tau l \; ,
\end{eqnarray}
the transformation  $\tilde{\delta}_{wgc}$  differs 
from ${\delta}_{r}$ by one more local symmetry,   ${\delta}_{h}(h(\tau))$, 
\begin{eqnarray}\label{f2}
 \delta_{h} \hat{Z}^M(\tau) & = 
h (\tau) (\hat{E}_{\tau}^\alpha E_\alpha^M(\hat{Z}) + 
\hat{\bar{E}}_{\tau}{}^{\dot\alpha} \bar{E}_{\dot\alpha}^M(\hat{Z}))\;	
\\ & \label{f2iE} \Leftrightarrow \quad \cases{ i_{h} \hat{E}^a =0 \; , \cr 
i_{h}\hat{E}^\alpha = h(\tau) \hat{E}_\tau ^\alpha\; , \cr 
i_{h} \hat{\bar{E}}{}^{\dot\alpha}= h(\tau) 
\hat{\bar{E}}_{\tau}^{\dot\alpha} \; , } \;   
\end{eqnarray}
namely, 
\begin{eqnarray}\label{rep=wgc}
{\delta}_{r}(r) = \tilde{\delta}_{wgc}(s=r) + {\delta}_{h}(h=-r) \; . 
\end{eqnarray}
Note that ${\delta}_{h}(h(\tau))$ is present in the 
Brink--Schwarz superparticle in any spacetime dimension $D$ where the 
gamma--matrices can be chosen to be symmetric.

\end{multicols}

\begin{thebibliography}{99}
\bibitem{Weyl} 
The relation between $D=4$ Majorana ($\Psi_{\underline{{\alpha}}}$)
and Weyl ($\Psi_\alpha, \bar{\Psi}_{\dot{\alpha}}= 
({\Psi}_{{\alpha}})^*$)
spinor notations are 
\begin{eqnarray}\label{4MsWs}
& \Psi_{\underline{{\alpha}}} \equiv 
\left(\matrix{\Psi_\alpha \cr \bar{\Psi}^{\dot{\alpha}}\cr}\right)\; , 
\quad \Psi_\alpha \equiv (\bar{\Psi}_{\dot{\alpha}})^*
\end{eqnarray} 
For $D=4$ 
fermionic supervielbein form  
$E^{\underline{{\alpha}}}$, 
\begin{eqnarray}
\label{4WsEal}
& E^{\underline{{\alpha}}} = 
\left(\matrix{E^\alpha, \bar{E}_{\dot{\alpha}}}\right)\; , \quad 
\end{eqnarray} 
and we can define 
\begin{eqnarray}\label{4MsEal1}
E_{\underline{{\alpha}}}
= C_{\underline{{\alpha}}\underline{{\beta}}}
E^{\underline{{\beta}}} =
\left(\matrix{E_\alpha \cr \bar{E}^{\dot{\alpha}}\cr}\right)
\; ,
\end{eqnarray} 
where 
\begin{eqnarray} 
\label{Cargec}
& 
C_{\underline{{\alpha}}\underline{{\beta}}}= 
\left(\matrix{ \epsilon_{\alpha\beta} & 0 \cr 
                   0 & \epsilon^{\dot{\alpha}\dot{\beta}} \cr     
}\right)\; , 
\\ \label{4WsEal2}
& E_\alpha=\epsilon_{\alpha\beta} E^{\beta} \; , \quad \bar{E}^{\dot{\alpha}}=
\epsilon^{\dot{\alpha}\dot{\beta}}\bar{E}_{\dot{\beta}} \; , 
\\ \nonumber &
\epsilon^{\alpha\gamma}\epsilon_{\gamma\beta}= \delta^\alpha_{\; \beta}\; , 
\quad  
\epsilon^{12}=1 =-\epsilon_{12}\; . 
\end{eqnarray}
The Majorana spinor (\ref{4WsEal}) satisfies 
\begin{eqnarray}\label{ThM}
(E^{\underline{{\beta}}})^\dagger 
{\cal B}_{\underline{{\beta}}^*}{}^{\underline{\alpha}} = 
E^{\underline{\alpha}}\; , \quad 
{\cal B}_{\underline{\beta}^*}{}^{\underline{{\alpha}}}
=  \left(\matrix{ 0        &    \delta^{\dot{\alpha}}{}_{\dot{\beta}} \cr 
               \delta_{\alpha}^{\beta} &    0 }\right)\; .
\end{eqnarray}
Note that, as a matrix, ${\cal B}= \gamma^0$, where 
\begin{eqnarray}
\label{gamma}
& \gamma_{\; \underline{{\alpha}}}^{a \; \underline{{\beta}}}= 
\left(\matrix{ 0 & \sigma^a_{\alpha\dot{\beta}} \cr 
       \tilde{\sigma}^{a\dot{\alpha}\beta} & 0 \cr }\right) \; .  \quad 
\end{eqnarray}
\end{thebibliography}
\end{document}